\documentclass[10pt,aps,prd,showkeys,nofootinbib,floatfix,amsmath,amssymb,showpacs,superscriptaddress,notitlepage]{revtex4-1}

\usepackage{rotating}
\usepackage{graphicx}
\usepackage{dcolumn}
\usepackage{bm}
\usepackage{color}
\usepackage{multirow}
\usepackage{mathtools}
\usepackage[caption=false]{subfig}
\usepackage{xr-hyper}
\usepackage[pdftex,bookmarks=true]{hyperref}
\hypersetup{
   colorlinks,
   citecolor=black,
   filecolor=black,
   linkcolor=black,
   urlcolor=black
}
\usepackage{cleveref}

\newcommand{\barc}{{\cal{B}}}
\newcommand{\barcs}{{\cal{B}^\ast}}
\newcommand{\bgb}{\cal{B} \gamma \rightarrow  \cal{B}^\ast  }
\newcommand{\bjb}{\cal{B}^{\ast} \cal{J}^\mu \cal{B}}

\newcommand{\occtooccs}{\Omega_{cc}^+ \gamma \to \Omega_{cc}^{\ast+}}
\newcommand{\xcctoxccs}{\Xi_{cc} \gamma \to {\Xi_{cc}^\ast}}
\newcommand{\xcctoxccsP}{\Xi_{cc}^{+} \gamma \to \Xi_{cc}^{\ast+}}
\newcommand{\xcctoxccsPP}{\Xi_{cc}^{++} \gamma \to \Xi_{cc}^{\ast++}}
\begin{document}

\title{Radiative transitions of doubly charmed baryons in lattice QCD}

\author{H. Bahtiyar}
\affiliation{Department of Physics, Mimar Sinan Fine Arts University, Bomonti 34380 Istanbul Turkey}
\author{K. U. Can}
\affiliation{RIKEN Nishina Center, RIKEN, Saitama 351-0198, Japan}
\author{G. Erkol}
\affiliation{Department of Natural and Mathematical Sciences, Faculty of Engineering, Ozyegin University, Nisantepe Mah. Orman Sok. No:34-36, Alemdag 34794 Cekmekoy, Istanbul Turkey}
\author{M. Oka}%
\affiliation{Advanced Science Research Center, Japan Atomic Energy Agency, Tokai, Ibaraki, 319-1195 Japan}
\author{T. T. Takahashi}%
\affiliation{Gunma National College of Technology, Maebashi, Gunma 371-8530 Japan}

\date{\today}

\begin{abstract}
We evaluate the spin-$3/2 \to$ spin-$1/2$ electromagnetic transitions of the doubly charmed baryons on 2+1 flavor, $32^3 \times 64$ PACS-CS lattices with a pion mass of $156(9)$ MeV/c$^2$. A relativistic heavy quark action is employed to minimize the associated systematic errors on charm-quark observables. We extract the magnetic dipole, $M1$, and the electric quadrupole, $E2$, transition form factors. In order to make a reliable estimate of the $M1$ form factor, we carry out an analysis by including the effect of excited-state contributions. We find that the $M1$ transition is dominant and light degrees of freedom ($u/d$- or $s$-quark) play the leading role. $E2$ form factors, on the other hand, are found to be negligibly small, which in turn, have minimal effect on the helicity and transition amplitudes. We predict the decay widths and lifetimes of $\Xi_{cc}^{\ast +,++}$ and $\Omega_{cc}^{\ast +}$ based on our results. Finite size effects on these ensembles are expected to be around 1\%. Differences in kinematical and dynamical factors with respect to the $N\gamma\to\Delta$ transition are discussed and compared to non-lattice determinations as well keeping possible systematic artifacts in mind. A comparison to $\Omega_c \gamma \rightarrow \Omega_c^\ast$ transition and a discussion on systematic errors related to the choice of heavy quark action are also given. Results we present here are particularly suggestive for experimental facilities such as LHCb, PANDA, Belle II and BESIII to search for further states. 
\end{abstract}
\pacs{14.20.Lq, 12.38.Gc, 13.40.Gp }
\keywords{charmed baryons, electric and magnetic form factor, lattice QCD}
\maketitle

\section{Introduction}
Recently there has been a profound interest in the spectroscopy and the structure of charmed baryons. Even though there are many states yet to be confirmed and discovered by experiments, charmed baryon sector holds its theoretical appeal. Binding of two heavy quarks and a light quark provides a unique view for confinement dynamics. All of the singly charmed ground-state baryons, which were predicted by the quark model, have been experimentally observed~\cite{Calicchio:1980sc,Biagi:1983en,Avery:1988uh,Jessop:1998wt,Biagi:1984mu}. Observation of the doubly-charmed baryons, on the other hand, have been challenging for experiments. First observation of the doubly charmed baryon was reported by SELEX collaboration in 2002 \cite{Mattson:2002vu}. Mass of the $\Xi_{cc}^{+}$ (ccd) baryon was reported as $3519 \pm 1$ MeV/c$^2$. However, none of the following experiments could confirm this result~\cite{Ratti:2003ez, Aubert:2006qw, Chistov:2006zj, Aaij:2013voa}, until very recently LHCb Collaboration discovered the isospin partner of $\Xi_{cc}^{+}$, namely $\Xi_{cc}^{++}$~\cite{Aaij:2017ueg}, containing two $c$ quarks and one $u$ quark. Mass of $\Xi_{cc}^{++}$ reported by LHCb is $3621.40 \pm 0.72 \pm 0.27 \pm 0.14$ MeV/c$^2$, approximately $100$ MeV larger than the SELEX finding and in agreement with lattice QCD predictions. This mass difference between the two isospin partners has been discussed with various theoretical approaches~\cite{Yao:2018ifh, Lu:2017meb, Ebert:2002ig, Karliner:2014gca}.

Spin-$1/2$ doubly-charmed baryons sit at the top layer of the flavor-mixed symmetric 20-plet of the SU(4) multiplet. In this layer, $\Xi_{cc}^{++}$ and $\Xi_{cc}^{+}$ are the isospin doublets, $I=1/2$, and $\Omega_{cc}$ is the isospin singlet, $I=0$. Spin-$3/2$ doubly-charmed baryons $\Xi_{cc}^{\ast++}$, $\Xi_{cc}^{\ast+}$ and $\Omega_{cc}^\ast$ sit at the third layer of the flavor-symmetric 20-plet with the same isospin assignments.

Electromagnetic properties of the baryon transitions give information about their internal structures and shape deformations. Examining the radiative transitions of doubly charmed baryons is a crucial element of understanding the heavy-quark dynamics. In our previous works, we have studied the $\Omega_c \gamma \rightarrow\Omega_c^\ast$ and $\Xi_c \gamma \rightarrow\Xi^\prime_c$ transitions in lattice QCD~\cite{Bahtiyar:2015sga, Bahtiyar:2016dom}. Being motivated by the recent experimental discovery of the $\Xi_{cc}^{++}$ baryon, we extend our investigations to the spin-$3/2 \to$ spin-$1/2$ electromagnetic transitions of the doubly charmed baryons. Such transitions are of particular interest for experimental facilities such as LHCb, PANDA, Belle II and BESIII to search for further states.

Spin-$3/2 \to$ spin-$1/2$ transitions are governed by three transition form factors, namely, the magnetic dipole ($M1$), the electric quadrupole ($E2$) and the electric charge quadrupole ($C2$). We study the Sachs form factors and the helicity amplitudes of these transitions and extract the decay width and the lifetime. Electromagnetic transitions of the doubly charmed baryons have also been studied within the heavy hadron chiral perturbation theory \cite{Meng:2017dni,Li:2017pxa,Li:2017cfz} and covariant baryon chiral perturbation theory \cite{Liu:2018euh}, in the context of bag model \cite{Hackman:1977am,Bernotas:2013eia} and quark models \cite{SilvestreBrac:1996bg,Lichtenberg:1976fi,JuliaDiaz:2004vh,Faessler:2006ft,Branz:2010pq,Oh:1991ws,Xiao:2017udy} and by QCD sum rules \cite{Sharma:2010vv,Cui:2017udv}. 

This paper is organized as follows: In \Cref{sec:lf}, we give the formulation of the transition kinematics. \Cref{sec:ls} presents the details of our lattice setup. We present and discuss our results in \Cref{sec:rd}, and summarize the work in \Cref{sec:sc}.

\section{Lattice Formulation}\label{sec:lf}
Electromagnetic transition form factors for a $\bgb$ process is encoded into baryon matrix elements written in the following form: 
\begin{equation}\label{eq:matel}
	\langle \barcs (p^\prime,s^\prime)|\mathcal{J}_\mu|\barc(p,s)\rangle= i \sqrt{\frac{2}{3}}\left(\frac{m_\barcs \
	m_\barc}{E_\barcs({\bf p^\prime})E_\barc({\bf p})}\right)\bar{u}_\tau(p^\prime,s^\prime) {\cal O}^{\tau\mu} u(p,s),
\end{equation}
where $\barc$ and $\barcs$ denote spin-$1/2$ and spin-$3/2$ baryons, respectively. $p$ and $p^\prime$ denote the initial and final four momenta and, $s$ and $s^\prime$ denote the spins. $u(p,s)$ is the Dirac spinor and $u_\tau(p,s)$ is the Rarita-Schwinger spin vector.
Operator $\cal{O}^{\tau\mu}$ can be parameterized in terms of Sachs form factors \cite{Jones:1972ky},
\begin{equation}\label{eq:sachs}
	{\cal O}^{\tau\mu}=G_{M1}(q^2) K_{M1}^{\tau\mu}+G_{E2}(q^2) K_{E2}^{\tau\mu}+G_{C2}(q^2) K_{C_2}^{\tau\mu},
\end{equation}
where $G_{M1}$, $G_{E2}$ and $G_{C2}$ denote the magnetic dipole, the electric quadrupole and the electric charge quadrupole transition form factors, respectively. The kinematical factors are defined as
\begin{align}
	K_{M1}^{\tau\mu}&=-3\Big((m_\barcs+m)^2-q^2\Big)^{-1}i\epsilon^{\tau\mu\alpha\nu}P^\alpha q^\nu~(m_\barcs+m_\barc)/2m_\barc,\\
	K_{E2}^{\tau\mu}&=-K_{M1}^{\tau\mu}-6\Omega^{-1}(q^2)~ i\epsilon^{\tau\beta\alpha\nu}P^\alpha q^\nu~ \epsilon^{\mu\beta\rho\theta}p^{\prime\rho} q^{\theta}~\gamma_5 (m_\barcs+m_\barc)/m_\barc,\\
	K_{C_2}^{\tau\mu}&=-3\Omega^{-1}(q^2)~q^\tau (q^2 P^\mu-q\cdot P~ q^\mu)~i\gamma_5(m_\barcs+m_\barc)/m_\barc.
\end{align}
Here $q=p^\prime-p$ is the transferred four\--momentum, $P=(p^\prime+p)/2$ and 
\begin{equation}
	\Omega(q^2)=\Big((m_\barcs+m_\barc)^2-q^2\Big)\Big((m_\barcs-m_\barc)^2-q^2\Big).
\end{equation}
The Rarita-Schwinger spin sum for the spin-$3/2$ field in Euclidean space is given by
\begin{align}
	\sum_s u_\sigma(p,s)& \bar{u}_\tau(p,s)=\frac{-i\gamma\cdot p+m_\barcs}{2m_\barcs}\left[g_{\sigma\tau}-\frac{1}{3}\gamma_\sigma \gamma_\tau+\frac{2p_\sigma p_\tau}{3m_\barcs^2}-i\frac{p_\sigma \gamma_\tau-p_\tau \gamma_\sigma}{3m_\barcs}\right],
\end{align}
and the Dirac spinor spin sum by
\begin{equation}
	\sum_s u(p,s) \bar{u}(p,s)=\frac{-i\gamma\cdot p +m_\barc}{2m_\barc}.
\end{equation}
To extract the form factors we use the following two- and three-point correlation functions,
 \begin{align}
 	\begin{split}\label{eq:deltacf}
 	&\langle G_{\sigma\tau}^{\barcs\barcs}(t; {\bf p};\Gamma_4)\rangle=\sum_{\bf x}e^{-i{\bf p}\cdot {\bf x}}\Gamma_4^{\alpha\alpha^\prime} \times \langle \text{vac} | T [\eta_{\sigma}^\alpha(x) \bar{\eta}_{\tau}^{\alpha^\prime}(0)] | \text{vac}\rangle,
 	\end{split}\\
 	\begin{split}\label{eq:nuccf}
 	&\langle G^{\barc\barc}(t; {\bf p};\Gamma_4)\rangle=\sum_{\bf x}e^{-i{\bf p}\cdot {\bf x}}\Gamma_4^{\alpha\alpha^\prime} \times \langle \text{vac} | T [\eta^\alpha(x) \bar{\eta}^{\alpha^\prime}(0)] | \text{vac}\rangle,
 	\end{split}\\
 	\begin{split}\label{eq:thrpcf}
 	&\langle G_\sigma^{\bjb}(t_2,t_1; {\bf p}^\prime, {\bf p};\mathbf{\Gamma})\rangle=-i\sum_{{\bf x_2},{\bf x_1}} e^{-i{\bf p}\cdot {\bf x_2}} e^{i{\bf q}\cdot {\bf x_1}} \mathbf{\Gamma}^{\alpha\alpha^\prime} \langle \text{vac} | T [\eta_\sigma^\alpha(x_2) j_\mu(x_1) \bar{\eta}^{\alpha^\prime}(0)] | \text{vac}\rangle,
 	\end{split}
 \end{align}
where the spin projection matrices are given as 
\begin{equation}
	\Gamma_i=\frac{1}{2}\left(\begin{matrix}\sigma_i & 0 \\ 0 & 0 \end{matrix}\right), \qquad \Gamma_4=\frac{1}{2}\left(\begin{matrix}I & 0 \\ 0 & 0 \end{matrix}\right).
\end{equation}
Here, $\alpha$, $\alpha^\prime$ are the Dirac indices, $\sigma$ and $\tau$ are the Lorentz indices of the spin-3/2 interpolating field and $\sigma_i$ are the Pauli spin matrices. Spin-$1/2$ state is created at $t=0$ and it interacts with the external electromagnetic field at time $t_1$ while it propagates to fixed-time $t_2$ where the final spin-$3/2$ state is annihilated.

We choose the interpolating fields similarly to those of $\Delta$ and $N$ as
\begin{align}
		&\eta_\mu(x)=\frac{1}{\sqrt{3}}\epsilon^{ijk} \left\{2[c^{T i}(x) C \gamma_\mu \ell^j(x)]c^k(x)+[c^{T i}(x) C \gamma_\mu c^j(x)]\ell^k(x)\right\},\label{eq:deltaint}\\
		&\eta(x)=\epsilon^{ijk}[c^{T i}(x) C \gamma_5 \ell^j(x)]c^k(x),
\end{align}
where $c$ denotes charm quark and $i$, $j$, $k$ are the color indices. Since we study the $\Xi_{cc}^{++}, \Xi_{cc}^{+}$ and $\Omega_{cc}^{+}$ baryons, $\ell$ is selected as $u$, $d$ and $s$ quark, respectively. Charge conjugation matrix is defined as $C=\gamma_4\gamma_2$. Interpolating field in \Cref{eq:deltaint} has been shown to have minimal overlap with spin-1/2 states and therefore does not need any spin-3/2 projection~\cite{Alexandrou:2014sha}.

To extract the form factors, we calculate the following ratio of the two- and three-point functions:
\begin{equation}\label{eq:ratio}
	R_\sigma(t_2,t_1;{\bf p}^\prime,{\bf p};\mathbf{\Gamma};\mu)=
\cfrac{\langle G_\sigma^{\bjb}(t_2,t_1; {\bf p}^\prime, {\bf p};\mathbf{\Gamma})\rangle}{\langle \delta_{ij} G_{ij}^{ \barcs \barcs}(t_2; {\bf p}^\prime;\Gamma_4)\rangle} \left[\cfrac{ \delta_{ij} G_{ij}^{ \barcs \barcs} (2t_1; {\bf p}^\prime;\Gamma_4)\rangle }{ G^{ \barc \barc} (2t_1; {\bf p};\Gamma_4)\rangle }\right]^{1/2}.
\end{equation} 
In the large Euclidean time limit, $t_2-t_1\gg a$ and $t_1\gg a$, time dependence of the correlators are eliminated so that the ratio in \Cref{eq:ratio} reduces to the desired form
\begin{equation}\label{eq:desratio}
	R_\sigma(t_2,t_1;{\bf p^\prime},{\bf p};\Gamma;\mu)\xrightarrow[t_2-t_1\gg a]{t_1\gg a} \Pi_\sigma({\bf p^\prime},{\bf p};\Gamma;\mu).
\end{equation}
We choose the ratio in \Cref{eq:ratio} from among several other alternatives~\cite{Leinweber:1992pv, Alexandrou:2003ea, Alexandrou:2004xn, Alexandrou:2007dt} as it leads to a good plateau region and signal quality~\cite{Bahtiyar:2015sga}.

Sachs form factors can be singled out choosing appropriate combinations of Lorentz direction $\mu$ and projection matrices $\Gamma$. Similar to our work in Ref.~\cite{Bahtiyar:2015sga}, we fix the kinematics for $\bgb$ (spin-$3/2$ at rest) as
\begin{align}
	\begin{split}\label{eq:cff}
	G_{C2}(q^2)&=C(\mathbf{q}^2)\frac{2m_{\cal{B}^\ast}}{\mathbf{q}^2} \Pi_k({\bf q},{\bf 0};i \Gamma_k;4)
	\end{split}
	\\
	\begin{split}\label{eq:mff}
	G_{M1}(q^2)&=C(\mathbf{q}^2)\frac{1}{|\mathbf{q}|}\left[\Pi_l(q_k,{\bf 0}; \Gamma_k;l)-\frac{m_{\cal{B}^\ast}}{E_{\cal{B}^\ast}}\Pi_k(q_k,{\bf 0}; \Gamma_l;l)\right], 
	\end{split}\\
	\begin{split}\label{eq:qff}
	G_{E2}(q^2)&=C(\mathbf{q}^2)\frac{1}{|\mathbf{q}|}\left[\Pi_l(q_k,{\bf 0}; \Gamma_k;l)+\frac{m_{\cal{B}^\ast}}{E_{\cal{B}^\ast}}\Pi_k(q_k,{\bf 0}; \Gamma_l;l)\right],  
	\end{split}
\end{align}
where
\begin{equation}
	C(\mathbf{q^2})= 2\sqrt{6} \frac{E_{\cal{B}} m_{\cal{B}}}{m_{\cal{B}^\ast}+m_{\cal{B}}}\left( 1+\frac{m_{\cal{B}}}{E_{\cal{B}}}\right)^{1/2} \left(1+\frac{\mathbf{q}^2}{3m_{\cal{B}^\ast}^2} \right)^{1/2}.
\end{equation}
Here, $k$ and $l$ are two distinct indices running from 1 to 3. For real photons, only $G_{M1}$ and $G_{E2}$ contribute. $G_{C2}$ does not play any role since it is proportional to the longitudinal helicity amplitude. In this work, we focus on the $M1$ and $E2$ transition form factors only due to poor signal-to-noise ratio of the $C2$ form factor with a limited number of gauge configurations.

\section{Lattice Setup}\label{sec:ls}

\subsection{Gauge Configurations}
We have run our simulations on gauge configurations generated by the PACS-CS collaboration~\cite{Aoki:2008sm} with the non-perturbatively $O(a)$-improved Wilson quark action and the Iwasaki gauge action. Details of the gauge configurations are given in \Cref{tab:gauge}. Simulations are carried out with near physical $u$,$d$ sea quarks of hopping parameter $\kappa^{\text{sea}}_{ud}=$ 0.13781. This corresponds to a pion mass of approximately 156~MeV~\cite{Aoki:2008sm}. Hopping parameter for the sea $s$ quark
is fixed to $\kappa^{\text{sea}}_{s} = 0.13640$.
\begin{table}[!htb]
\centering
	\setlength{\extrarowheight}{5pt}
\caption{Details of the gauge configurations that we employ~\cite{Aoki:2008sm}.  We list the spatial and temporal sizes of the lattice ($N_s$ and $N_t$), number of flavors ($N_f$), the lattice spacing ($a$) and inverse lattice spacing ($a^{-1}$), spatial extent of the lattice ($L$), inverse gauge coupling ($\beta$), Clover coefficient ($c_{sw}$), hopping parameter of the quark with flavor $f$ ($\kappa^{\text{sea}}_f$) and the corresponding pion mass ($m_\pi$). We make our measurements on 163 and 194 configurations, respectively for $\Xi_{cc}$ and $\Omega_{cc}$. }
\begin{tabular*}{\textwidth}{@{\extracolsep{\fill}}cccccccccc}
	\hline \hline
 	$N_s \times N_t$ & $N_f$  &  $a$ [fm] & $a^{-1}$ [GeV] & $L$ [fm] & $\beta$ & $c_{sw}$& $\kappa^{\text{sea}}_{ud}$  & $\kappa^{\text{sea}}_s$  & $m_\pi$ [MeV] \\
 	\hline
 	{$32^3 \times 64$} & $2+1$ & $0.0907(13)$ & $2.176(31)$ & $2.90$ & $1.90$ & $1.715$ & $0.13781$ & $0.13640$ & $156(7)(2)$  \\ 
 	\hline \hline
\end{tabular*}
\label{tab:gauge}
\end{table}
It has been shown that it is feasible to carry-out simulations involving charm quarks on ensembles with physical light dynamical quarks~\cite{PhysRevD.84.074505}. Since the ensemble we employ has almost-physical quark masses, we omit an extrapolation to the physical light quark mass point. A comparison of our previous $m_{\Omega_c}$ results from Ref.~\cite{Can:2013tna} (extrapolated value: $2.740(24)$ GeV) and Ref.~\cite{Bahtiyar:2015sga} (this ensemble: $2.750(15)$ GeV) along with a more recent $\chi$PT form extrapolation on $m_{\Sigma_c}$ (extrapolated: $2.487(31)$ GeV vs. this ensemble: $2.486(47)$ GeV) from Ref.~\cite{Can:2017pee} indicates that almost-physical ensemble values agree with extrapolated results. Therefore, we consider the extracted values on this ensemble as final, which eliminates one source of systematic error.  

\subsection{Strange quark mass re-tuning}\label{sec:sq}

\begin{figure}[htb]
	\centering
	\includegraphics[width=\textwidth]{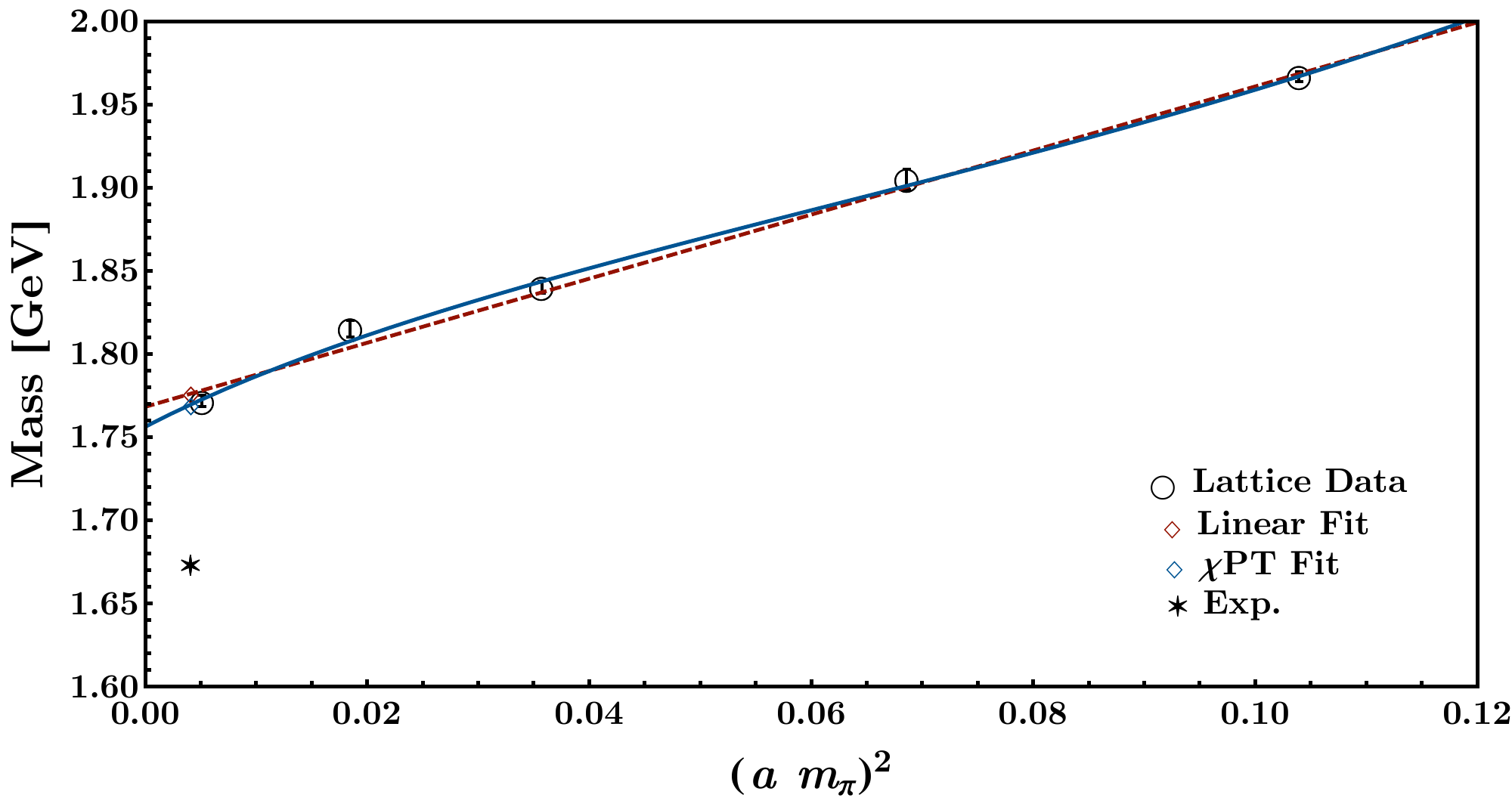}
	\caption{$m_\pi^2$ dependence of $m_\Omega$ values. Black lattice data points are taken from Ref.~\cite{Aoki:2008sm}. Red curve is a linear, $a + b m_\pi^2$, fit form while the blue curve is Equation (15) of Ref.~\cite{Tiburzi:2008bk}. Empty diamonds show the extrapolated values and the black star is the experimental $m_\Omega$.}
	\label{fig:omegachpt}
\end{figure}

We have been unable to reproduce the experimental $\Omega$ mass in our previous studies with $\kappa_s = 0.13640$ as tuned by the PACS-CS Collaboration to physical strange quark mass with respect to the mass of $\Omega$ baryon. Our determination of the mass of $\Omega$ on the $\kappa^{\text{sea}}_{ud}=0.13781$ ensemble with $\kappa_s^{\text{val}} = 0.13640$ is $m_\Omega = 1.790(17)$ GeV, which overestimates the experimental value by $\sim 6\%$~\cite{Can:2015exa}. It is, however, in agreement with the PACS-CS value reported from the same ensemble, $m_\Omega = 1.772(7)$ GeV~\cite{Aoki:2008sm}. A crude analysis of the $m_\Omega$ values reported by PACS-CS is shown in \Cref{fig:omegachpt}. We employ a linear and a $\chi$PT form~\cite{Tiburzi:2008bk} for extrapolation, both of which overestimate the experimental value. This issue with the tuning of $\kappa_s$ has been raised in some works in the literature as well~\cite{PhysRevLett.108.112001,Mohler:2013rwa}. Therefore we opt-in to use a partially quenched strange quark $m_s^{\text{val}} \ne m_s^{\text{sea}}$ and adopt the value $\kappa_s^{\text{val}} = 0.13665$ reported in Ref.~\cite{PhysRevLett.108.112001} while keeping $a^{-1}=2.176(31)$ GeV. We find $m_\Omega = 1.674(30)$ GeV with the re-tuned $\kappa_s$ value. 

\subsection{Heavy quark action and quark mass tuning}\label{sec:tsu}
\begin{figure}[htb]
	\centering
	\includegraphics[width=\textwidth]{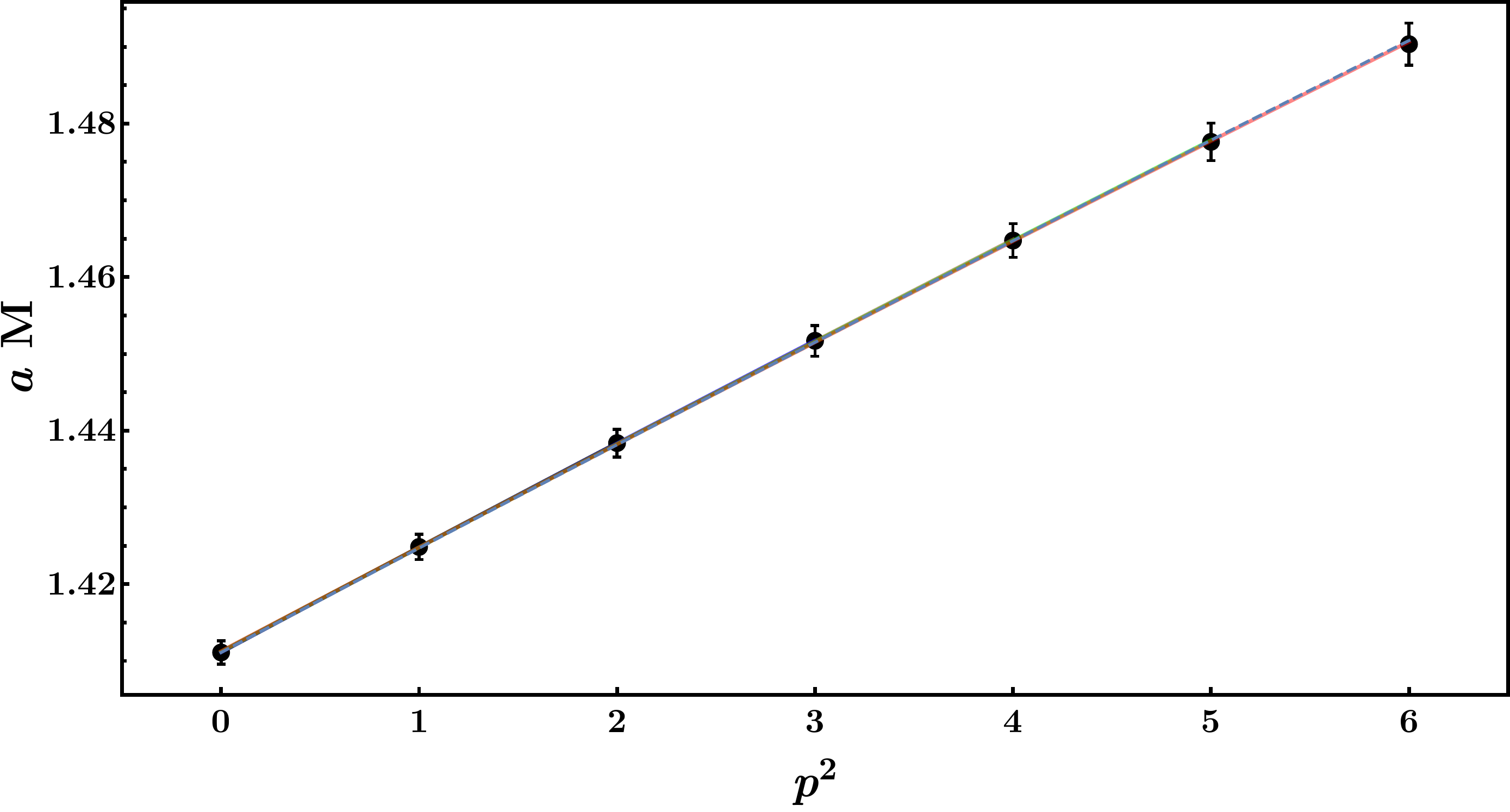}
	\caption{Relativistic dispersion relation of the 1S charmonium state. Black data points are $E_{1S}(\mathbf{p})$ extracted from fits to \Cref{eq:2ptf}. Lines show the fits to \Cref{eq:disp} where $c_{\text{eff}}^2$ is considered as a free parameter. Barely visible dashed blue line is \Cref{eq:disp} with $c_{\text{eff}}^2=1$.}
	\label{fig:1s_disp}
\end{figure}

It is well known that the Clover action has $\mathcal{O}(m_Q a)$ discretization errors that might become significant for charm quarks. Although we have successfully utilized the Clover action for charm quarks in our previous works while accounting for the associated errors, in this work we improve our simulations with a relativistic heavy quark action. We employ the so-called \emph{Tsukuba action}, proposed by Aoki et al.~\cite{PTP.109.383}, which is designed to remove the leading cutoff effects of order $(m_Q a)^n$ and reduce it to $\mathcal{O}(f(m_Q a)(a \Lambda_{QCD})^2)$ where $f(m_Q a)$ is an analytic function around the $m_Q a=0$ point and can be removed further by tuning the parameters of the action non-perturbatively. As a result, only $\mathcal{O}((a \Lambda_{QCD})^2)$ discretization errors remain. The action is
\begin{equation}
	S_\Psi = \sum_{x,y} \bar{\Psi}_x D_{x,y} \Psi_y,  	
\end{equation}   
where $\Psi$s are the heavy quark spinors and the fermion matrix is given as
\begin{align}
	\begin{split}
		D_{x,y} = \delta_{xy} 	
		&- \kappa_Q \sum_{\mu=1}^3 \left[ (r_s - \nu \gamma_\mu) U_{x,\mu} \delta_{x+\hat\mu,y} + (r_s + \nu \gamma_\mu) U^\dag_{x,\mu} \delta_{x,y+\hat\mu} \right] \\
		&- \kappa_Q \left[ (1 - \gamma_4) U_{x,4} \delta_{x+\hat4,y} + (1 + \gamma_4) U^\dag_{x,4} \delta_{x,y+\hat4} \right] \\ 
		&- \kappa_Q \left[ c_B \sum_{\mu,\nu} F_{\mu\nu}(x) \sigma_{\mu\nu} + c_E \sum_{\mu} F_{\mu 4}(x) \sigma_{\mu 4} \right] \delta_{xy}.
	\end{split}
\end{align}
Here, the parameters $r_s$, $\nu$, $c_B$ and $c_E$ should be tuned in order to remove the discretization errors appropriately. We adopt the perturbative estimates for $r_s$, $c_B$ and $c_E$~\cite{Aoki:2004271} and non-perturbatively tuned $\nu$ value~\cite{Namekawa:2013vu}. We re-tune $\kappa_Q$ non-perturbatively so as to reproduce the relativistic dispersion relation,
\begin{equation}\label{eq:disp}
	E^2_{1S}(\mathbf{p}) = E^2_{1S}(\mathbf{0}) + c_{\text{eff}}^2 |\mathbf{p}|^2,
\end{equation}
for 1S spin-averaged charmonium state. We extract the energies of the pseudoscalar and vector charmonium states from the two-point correlation functions of the interpolating fields
\begin{equation}\label{eq:chint}
	\chi(x) = \bar{c} \gamma_5 c, \quad \chi_\mu(x) = \bar{c} \gamma_\mu c.
\end{equation}
The values of the parameters and extracted charmonium masses are given in \Cref{tab:tsu}. Masses of the charmonium states are in very good agreement with the experimental results. We give the extracted static masses, $E^2_{1S}(\mathbf{0})$, and effective speed of light, $c_{\text{eff}}^2$, in \Cref{tab:disp} and \Cref{fig:1s_disp} shows the dispersion relation. Hyperfine splitting is a simple prediction one can get from this exercise and is also a good indicator for the severeness of the discretization errors. Experimental $V-PS$ hyperfine splitting is $\Delta E_{(V-PS)} = 113$ MeV where our results yield $\Delta E_{V-PS} = 116(4)$ MeV. We do not include disconnected diagrams in this calculation hence the effects of the possible annihilation of the $\eta_c$ and $J/\psi$ into light hadrons are neglected. This mechanism would mainly affect the $\eta_c$ meson and lead to a mass shift of $\Delta M_{\eta_c} = -3 $ MeV~\cite{Brown:2014ena}. Considering this correction, our hyperfine splitting estimate increases by $3$ MeV in good agreement with the experimental result.    
\begin{table}[!htb]
\centering
\setlength{\extrarowheight}{5pt}
\caption{Parameter values of the relativistic heavy quark action and masses of pseudoscalar, vector and 1S charmonium states as well as the $V-PS$ hyperfine splitting.}
	\begin{tabular*}{\textwidth}{@{\extracolsep{\fill}}ccccccccc}
		\hline \hline
		$\kappa_Q$ & $r_s$ & $\nu$ & $c_B$ & $c_E$ & $m_{\eta_c}$ [GeV] & $m_{J/\psi}$ [GeV] & $m_{1S}$ [GeV] & $\Delta E_{(V-PS)}$ [MeV] \\
		\hline
		$0.10954007$ & $1.1881607$ & $1.1450511$ & $1.9849139$ & $1.7819512$ & $2.984(2)$ & $3.099(4)$ & $3.071(4)$ & $116(4)$ \\ 
		\hline \hline
	\end{tabular*}
\label{tab:tsu}
\end{table}
\begin{table}[!htb]
\centering
\setlength{\extrarowheight}{2pt}
\caption{Extracted static masses, $E_{1S}(\mathbf{0})$, in lattice and physical units and effective speed of light, $c_{\text{eff}}^2$ from the dispersion relation analysis with different momenta. $|\mathbf{p}|^2$ column indicates the number of momentum units used for the analysis.}
	\begin{tabular*}{\textwidth}{@{\extracolsep{\fill}}cccc}
		\hline \hline
		$|\mathbf{p}|^2 $ & $E_{1S}(\mathbf{0})$ [a] & $E_{1S}(\mathbf{0})$ [GeV] & $c_{\text{eff}}^2$ \\
		\hline
		$2$ & $1.41111 \pm 0.00150591$ & $3.07058 \pm 0.00327686$ & $1.00818 \pm 0.0159342$\\
		$3$ & $1.41113 \pm 0.00150235$ & $3.07063 \pm 0.00326911$ & $1.00538 \pm 0.0169947$\\
		$4$ & $1.41117 \pm 0.00149903$ & $3.07071 \pm 0.00326189$ & $1.00186 \pm 0.0175885$\\
		$5$ & $1.41122 \pm 0.00149308$ & $3.07082 \pm 0.00324894$ & $0.998545 \pm 0.0185763$\\
		$6$ & $1.41127 \pm 0.00148551$ & $3.07092 \pm 0.00323247$ & $0.995832 \pm 0.0197037$\\
		\hline \hline
	\end{tabular*}
\label{tab:disp}{}
\end{table}

\subsection{Simulation Details}
We make our simulations at the lowest allowed lattice momentum transfer $q = 2\pi/L$, corresponding to three-momentum squared value of $q^2=0.183$~GeV$^2$, where $L = N_s\,a$ is the spatial extent of the lattice. We use a simple scaling method as in Ref.~\cite{Leinweber:1992pv} in order to estimate the values of the form factors at zero momentum. We assume that the momentum-transfer dependence of the transition form factors is the same as the momentum dependence of the $\Omega_{cc}^\ast$ and $\Xi_{cc}^\ast$ baryon's charge form factors. Such a scaling was used in previous analyses~\cite{Leinweber:1992pv} and also suggested by the experimental analysis of the proton form factors. The scaling method provides more precise determination of the form factor values at zero momentum since extrapolations in finite momentum have to build on a functional form that suffer from large statistical errors. With the aid of this simple scaling, $G_{M1}(0)$ is estimated by
\begin{equation}\label{eq:scaling}
	G_{M1}^{s,c}(0)=G_{M1}^{s,c}(q^2)\frac{G_{E0}^{s,c}(0)}{G_{E0}^{s,c}(q^2)}.
\end{equation} 
We consider quark contributions separately due to the fact that their charge form factor contributions scale differently. We have observed that ~\cite{Can:2013tna,Can:2013zpa} the light-quark contribution produces a soft form factor while that of the heavy quark is harder, which falls off more slowly with increasing momentum-transfer squared. Since we found out similar results for different kinematics in our previous works~\cite{Bahtiyar:2015sga}, we fix the kinematics to where the spin-$3/2$ baryon is produced at rest and the spin-$1/2$ has momentum \textbf{−-q}.

In order to increase statistics, we insert positive and negative momenta in all spatial directions and make a simultaneous fit over all data. We consider current insertion along all spatial directions. The source-sink time separation is fixed to 12 lattice units (1.09 fm), which has been shown to be enough to avoid excited-state contaminations for electromagnetic form factors of singly charmed baryons \cite{Can:2013tna}. We have computed various source-sink pairs by shifting them by $12 a$. We perform $880$ and $600$ measurements for the $\Omega_{cc}$ and $\Xi_{cc}$ system respectively and bin the data with a bin size of $20$ in order to account for autocorrelations. To study the excited state effects, we make calculations with $14a$ (1.27 fm) and $15a$ (1.36 fm) separations on a subset of the gauge configurations also. All statistical errors are estimated by a single-elimination jackknife analysis. The vector current we utilize in our simulations is the point-split lattice vector current
  \begin{equation}
		 j_\mu = \frac{1}{2} [\overline{q}(x+\mu) U_\mu^{\dagger}(1+\gamma_\mu)q(x)-\overline{q}(x)U_\mu(1-\gamma_\mu)q(x+\mu)],
		 \label{eq:pointsplit}
  \end{equation}
which is conserved by Wilson fermions, thus eliminates the need for renormalization.

In order to improve the ground-state coupling, non-wall smeared source and sink are smeared in a gauge-invariant manner using a Gaussian form. In the case of light and strange quarks, we choose the smearing parameters so as to give a rms radius of $r^{l,s}_{\text{rms}} \sim 0.5$ fm. We have measured the size of the charm quark charge radius to be small compared to the light and strange quarks, both in mesons~\cite{Can:2012tx} and baryons~\cite{Can:2013tna}. Therefore, we adjust the smearing parameters to obtain $\langle r^c_{\text{rms}} \rangle = \langle r^{l,s}_{\text{rms}} \rangle / 3$. We use wall-source/sink method \cite{Can:2012tx} which provides a simultaneous extraction of all spin, momentum and projection components of the correlators. The wall source/sink is a gauge-dependent object that requires fixing the gauge. We fix the gauge to Coulomb, which gives a somewhat better coupling to the ground state. Note that using different smearing operators on source and sink leads to different overlap factors hence different ground-state coupling characteristics. This is visible as an asymmetric signal in our case. 

The effects of disconnected diagrams are neglected in this work since they are noisy and costly to compute. Furthermore contributions of disconnected diagrams to isovector electromagnetic form factors are usually suppressed \cite{Alexandrou:2007xj}. We also expect the sea-quark effects to be suppressed in our results.

\section{Results And Discussion}\label{sec:rd}
\subsection{Baryon masses}\label{sec:bm}
\begin{figure}
	\includegraphics[width=\textwidth]{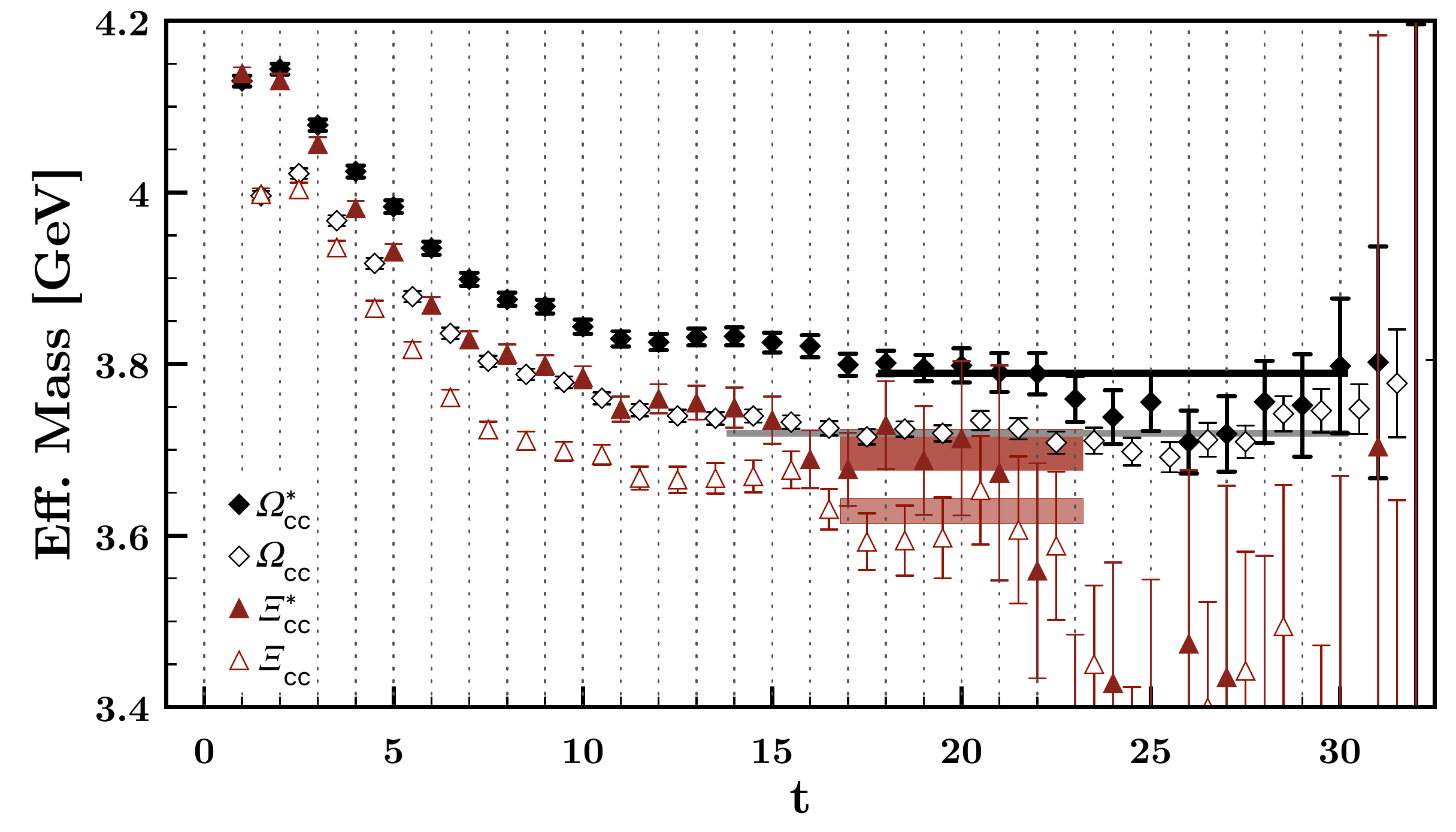}
	\caption{Effective mass plots for the doubly charmed baryons. Shaded bands show the fit regions. Empty symbols are slightly shifted to the right for a clearer view.}
	\label{fig:masses}
\end{figure}

\begin{figure}
	\includegraphics[width=\textwidth]{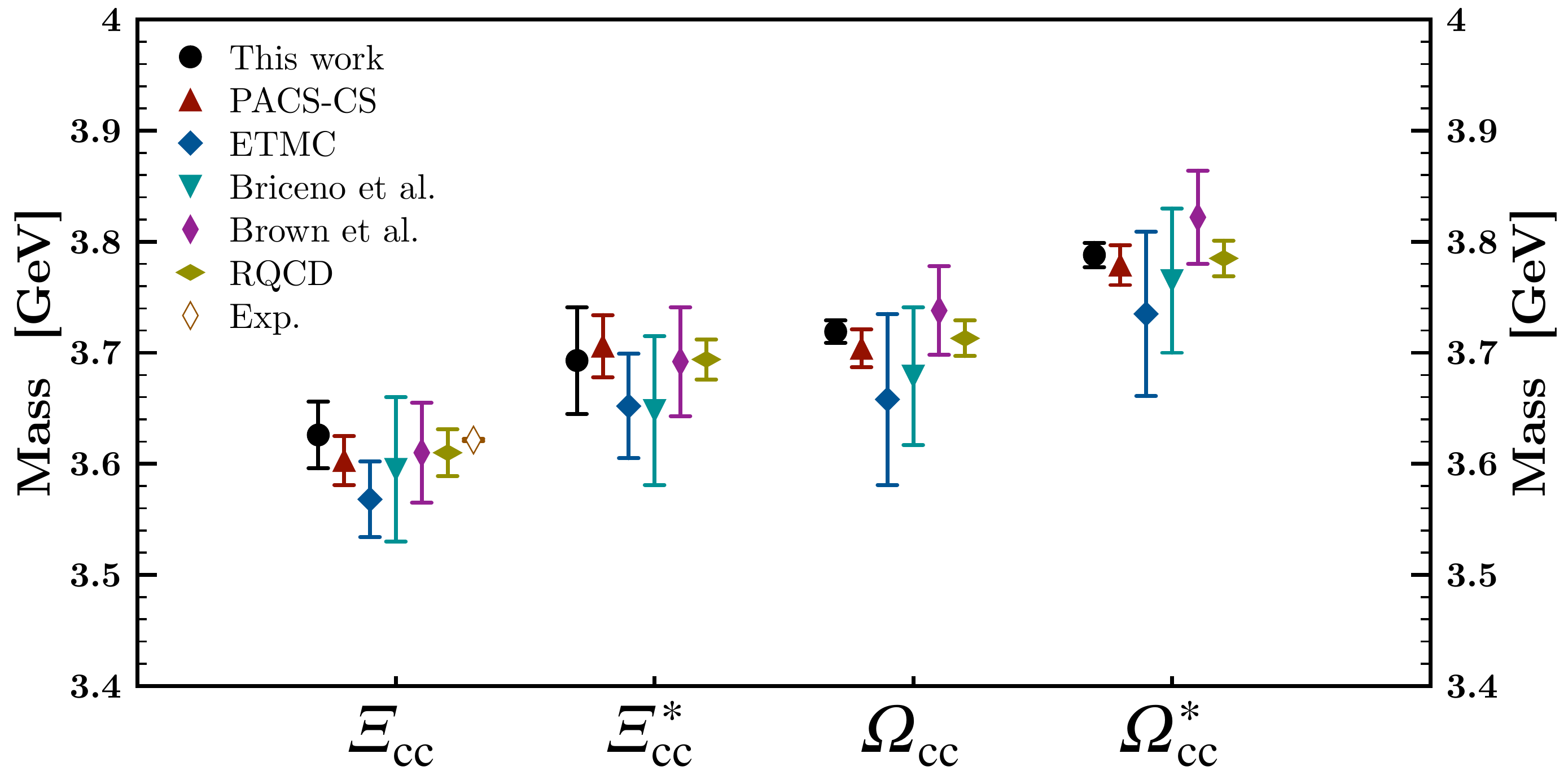}
	\caption{Visual comparison of our masses to other select lattice collaborations' results. Our errors are statistical only whereas other collaborations' are statistical and systematical errors combined. See \Cref{tab:mass} for references. }
	\label{fig:mass_comp}
\end{figure}

We extract the masses of spin-$1/2$ and spin-$3/2$ $\Omega_{cc}$ and $\Xi_{cc}$ baryons using their respective two-point correlation functions defined in \Cref{eq:nuccf,eq:deltacf}. In case of spin-$3/2$ baryons, an average over spatial Lorentz indices is taken. Two-point correlation functions reduce to
\begin{equation}\label{eq:2ptf}
	\langle G^{\barc \barc}(t; \mathbf{p}; \Gamma_4) \rangle \simeq Z_\barc(\mathbf{p}) \bar{Z}_\barc(\mathbf{p}) e^{-E_\barc(\mathbf{p}) t} (1 + \mathcal{O}(e^{-\Delta E t}) + \dots),
\end{equation}
where the mass of a baryon is encoded into the leading order exponential behavior and can be identified for the $\mathbf{p}=(0,0,0)$ case when the excited states are properly suppressed. We perform an effective mass analysis,
\begin{equation}
	m_{\rm eff}(t + \frac{1}{2}) = \text{ln} \frac{G^{\barc \barc}(t; \mathbf{0}; \Gamma_4)}{G^{\barc \barc}(t+1; \mathbf{0}; \Gamma_4)},	
\end{equation} 
 in order to estimate a suitable fit window, $[t_i, t_f]$, for the correlation functions and extract the masses by performing a non-linear regression analysis via \Cref{eq:2ptf}. It is possible to take the contributions of first excited states into account as correction terms to \Cref{eq:2ptf} to enhance the analysis, however, we find it to be an excessive treatment considering the precision and agreement of our results. Initial time slice $t_i$ is chosen by intuition where the data starts to form a plateau while the fit window is extended to the time slice until when the signal is deemed to be lost. Effective mass plots are shown in \Cref{fig:masses}. Fit regions are determined to be $[t_i, t_f] = [17,23]$, $[17,23]$, $[14,30]$ and $[18,30]$ for $\Xi_{cc}$, $\Xi_{cc}^\ast$, $\Omega_{cc}$ and $\Omega_{cc}^\ast$ baryons respectively. Our results are given in \Cref{tab:mass} and shown in \Cref{fig:mass_comp} in comparison to other determinations by various lattice collaborations and the experimental values where available. Note that our results are obtained at a pion mass of $m_\pi \approx$ 156 MeV and compare well to those from other lattice collaborations which are either on physical-quark mass point or extrapolated to physical quark mass and considers the continuum limit.
\begin{table}[ht]
\centering
\caption{Extracted $\Xi_{cc}$, $\Xi_{cc}^\ast$, $\Omega_{cc}$, and $\Omega_{cc}^\ast$ masses as well as those of other lattice collaborations and experimental values. The errors in this work are statistical only, while those quoted by other collaborations correspond to statistical and various systematical errors if given.}
\label{tab:mass}
	\setlength{\extrarowheight}{5pt}
	\begin{tabular*}{\textwidth}{@{\extracolsep{\fill}}cccccccc}
	\hline \hline
	& This work & PACS-CS \cite{Namekawa:2013vu} & ETMC \cite{Alexandrou:2014sha} & Briceno et al. \cite{Briceno:2012wt} & Brown et al. \cite{Brown:2014ena} & RQCD \cite{PhysRevD.92.034504}& Experiment \cite{Aaij:2017ueg}\\ \hline
	$m_{\Xi_{cc}}$ [GeV]			& 3.626(30)	& 3.603(22)	& 3.568(14)(19)(1)	& 3.595(39)(20)(6)	& 3.610(23)(22)		& 3.610(21) &3.62140(72)(27)(14) \\ 
	$m_{\Xi_{cc}^\ast}$ [GeV]		& 3.693(48)	& 3.706(28)	& 3.652(17)(27)(3)	& 3.648(42)(18)(7)	& 3.692(28)(21)     & 3.694(18) &--- \\	 
	$m_{\Omega_{cc}}$ [GeV]			& 3.719(10)	& 3.704(17)	& 3.658(11)(16)(50)	& 3.679(40)(17)(5)	& 3.738(20)(20)		& 3.713(16) &--- \\ 
	$m_{\Omega_{cc}^\ast}$ [GeV]	& 3.788(11)	& 3.779(18)	& 3.735(13)(18)(43)	& 3.765(43)(17)(5)	& 3.822(20)(22)		& 3.785(16) &--- \\ 
 	\hline \hline
 	\end{tabular*}
\end{table}
\subsection{Form factors}\label{sec:ff}
Since we have all possible Lorentz, momentum, polarization and current indices, we define an average over correlation function ratios,
\begin{equation}
\label{eq:pi1pi2}
	\Pi_1=\frac{C(q^2)}{|\bf{q}|}\frac{1}{6}\sum_{k,l}\Pi_l(q_k,{\bf 0}; \Gamma_k;l), \quad
	\Pi_2=\frac{C(q^2)}{|\bf{q}|}\frac{1}{6}\sum_{k,l}\Pi_k(q_k,{\bf 0}; \Gamma_l;l),
\end{equation}
and rewrite \Cref{eq:mff,eq:qff} as,
\begin{align}
	\begin{split}\label{eq:mffavg}
	G_{M1}(q^2)&=\Pi_1-\frac{m_{\cal{B}^\ast}}{E_{\cal{B}^\ast}}\Pi_2, 
	\end{split}\\
	\begin{split}\label{eq:qffavg}
	G_{E2}(q^2)&=\Pi_1+\frac{m_{\cal{B}^\ast}}{E_{\cal{B}^\ast}}\Pi_2.
	\end{split}
\end{align}
Note that the factor in front of the $\Pi_2$ term simplifies to $m_{\cal{B}^\ast}/E_{\cal{B}^\ast} = 1$ since we only calculate the kinematical case where the spin-$3/2$ particle is at rest. Also let us remind the reader that we omit the $C_2$ form factor due to its poor signal-to-noise ratio. 
\subsubsection{Excited-state contamination and multi-exponential fits}\label{sec:exct}
\begin{figure}[htb]
	\includegraphics[width=.49\textwidth]{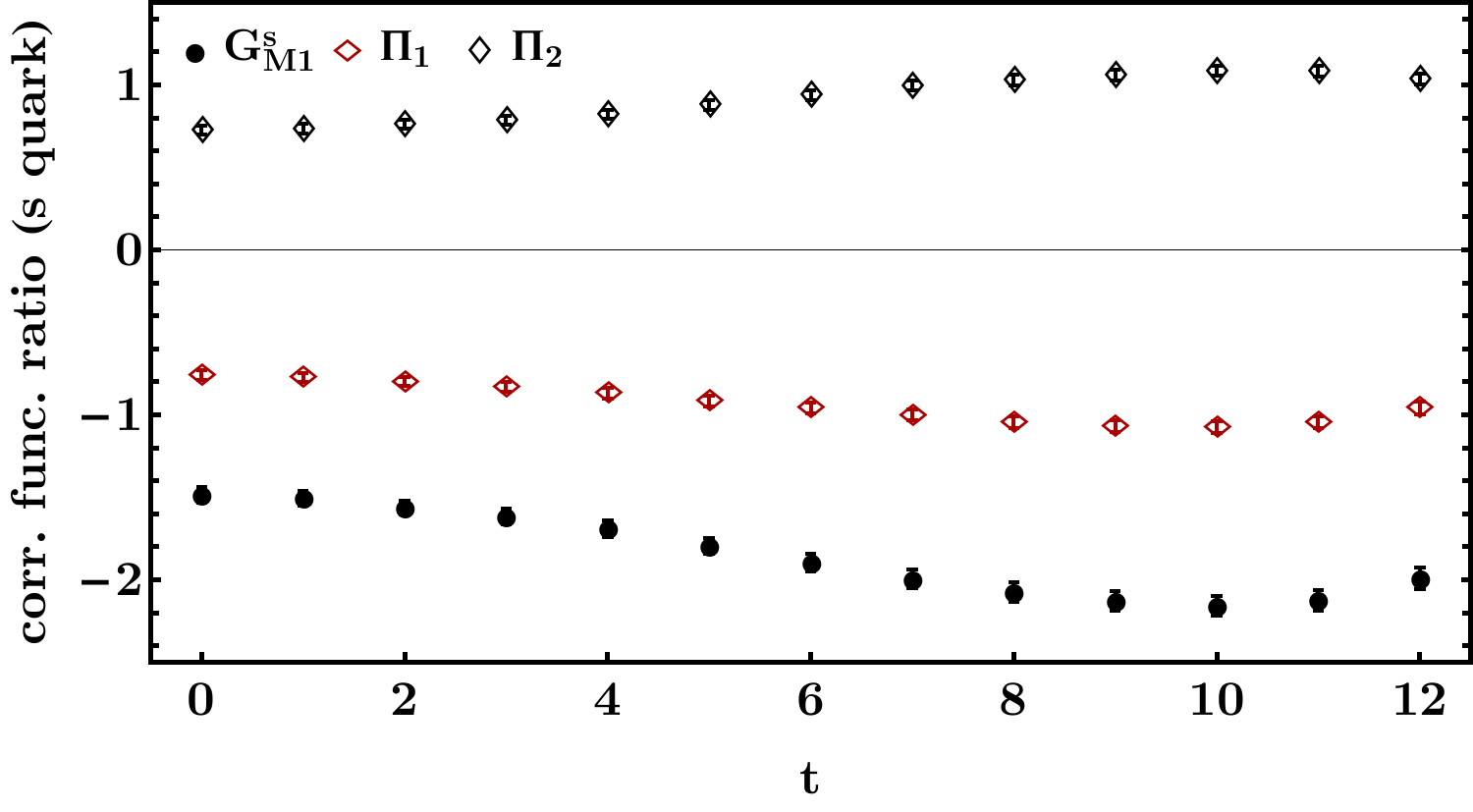} \includegraphics[width=.49\textwidth]{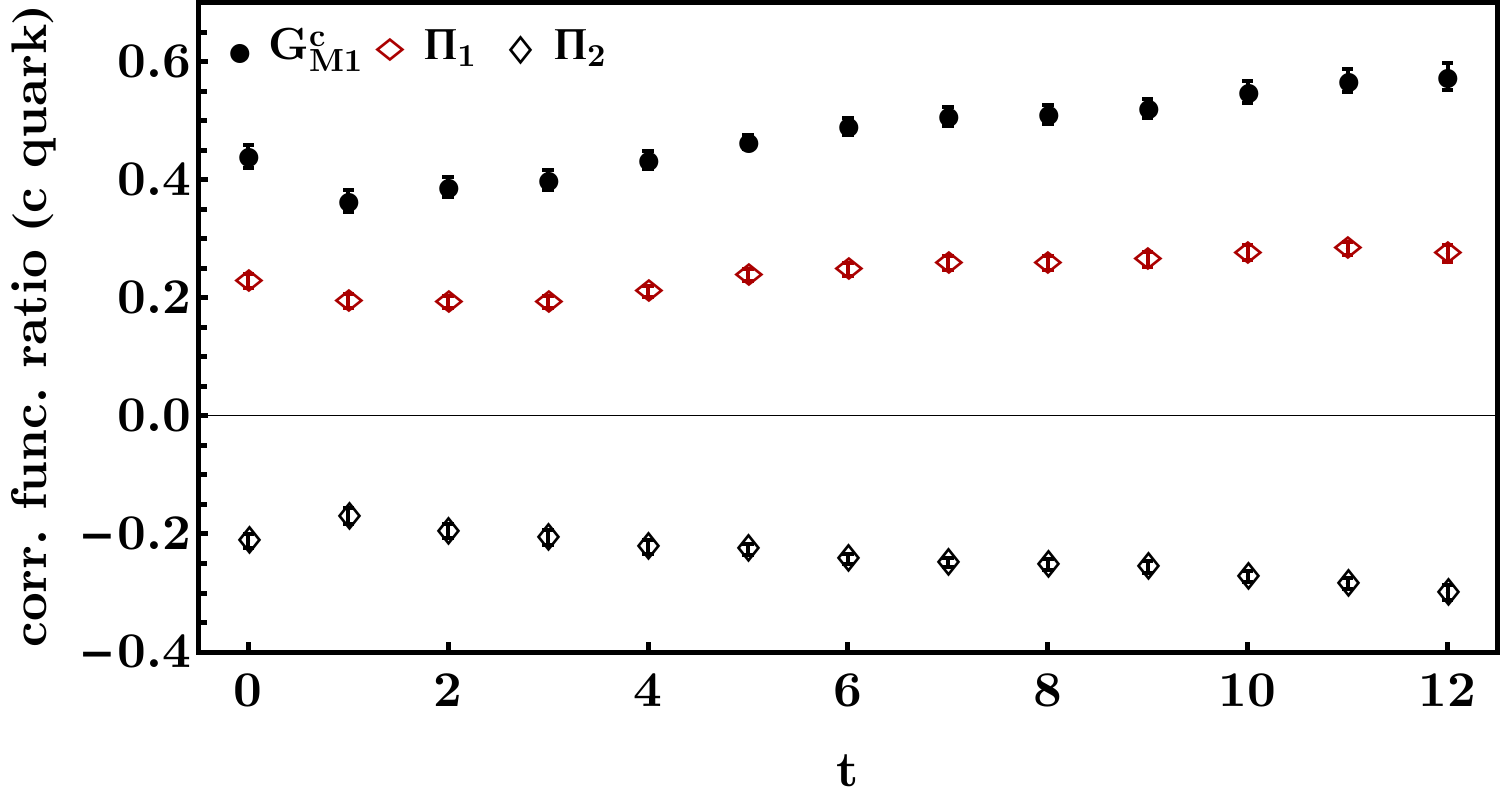} \\
	\includegraphics[width=.49\textwidth]{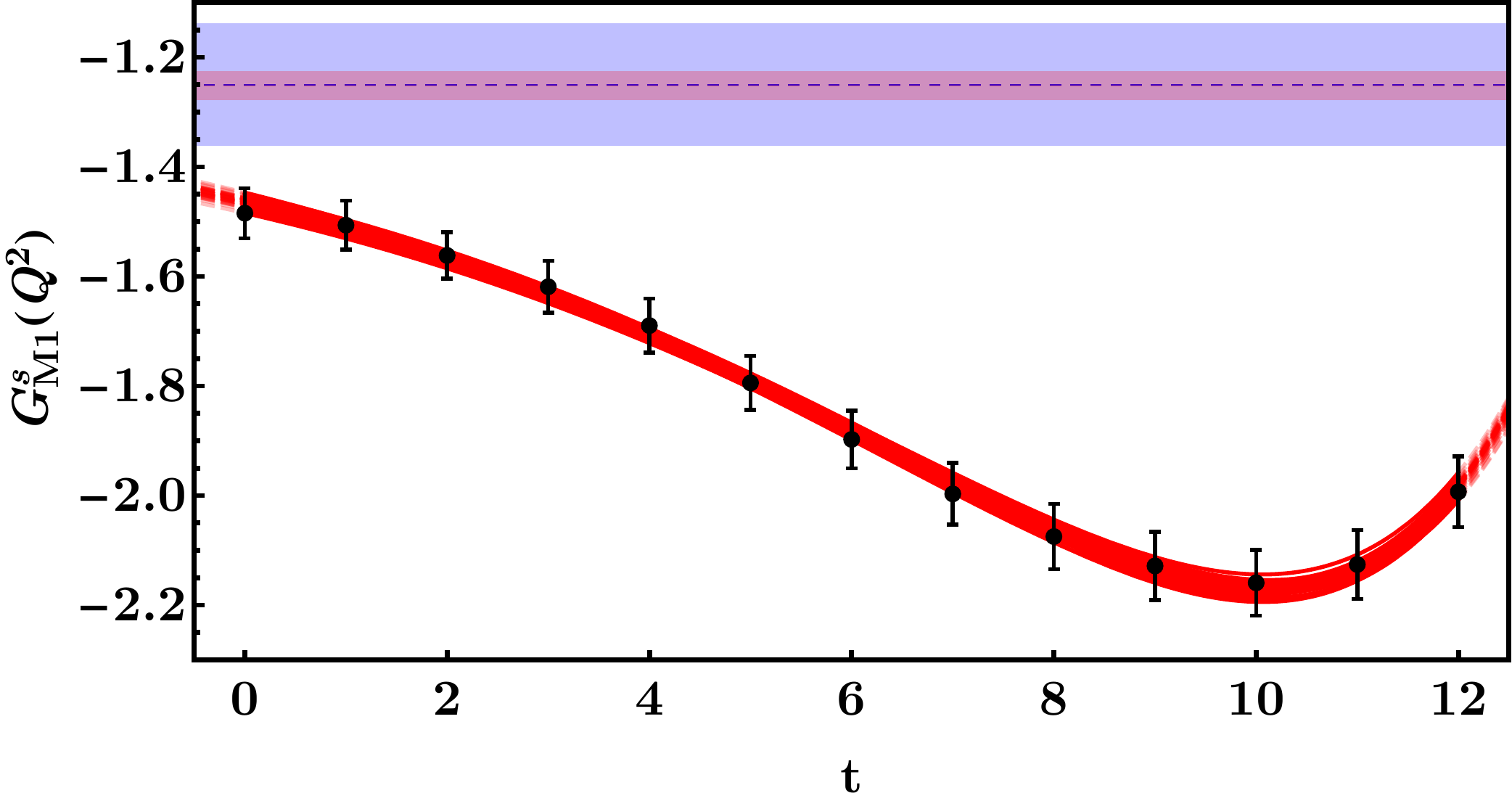} \includegraphics[width=.49\textwidth]{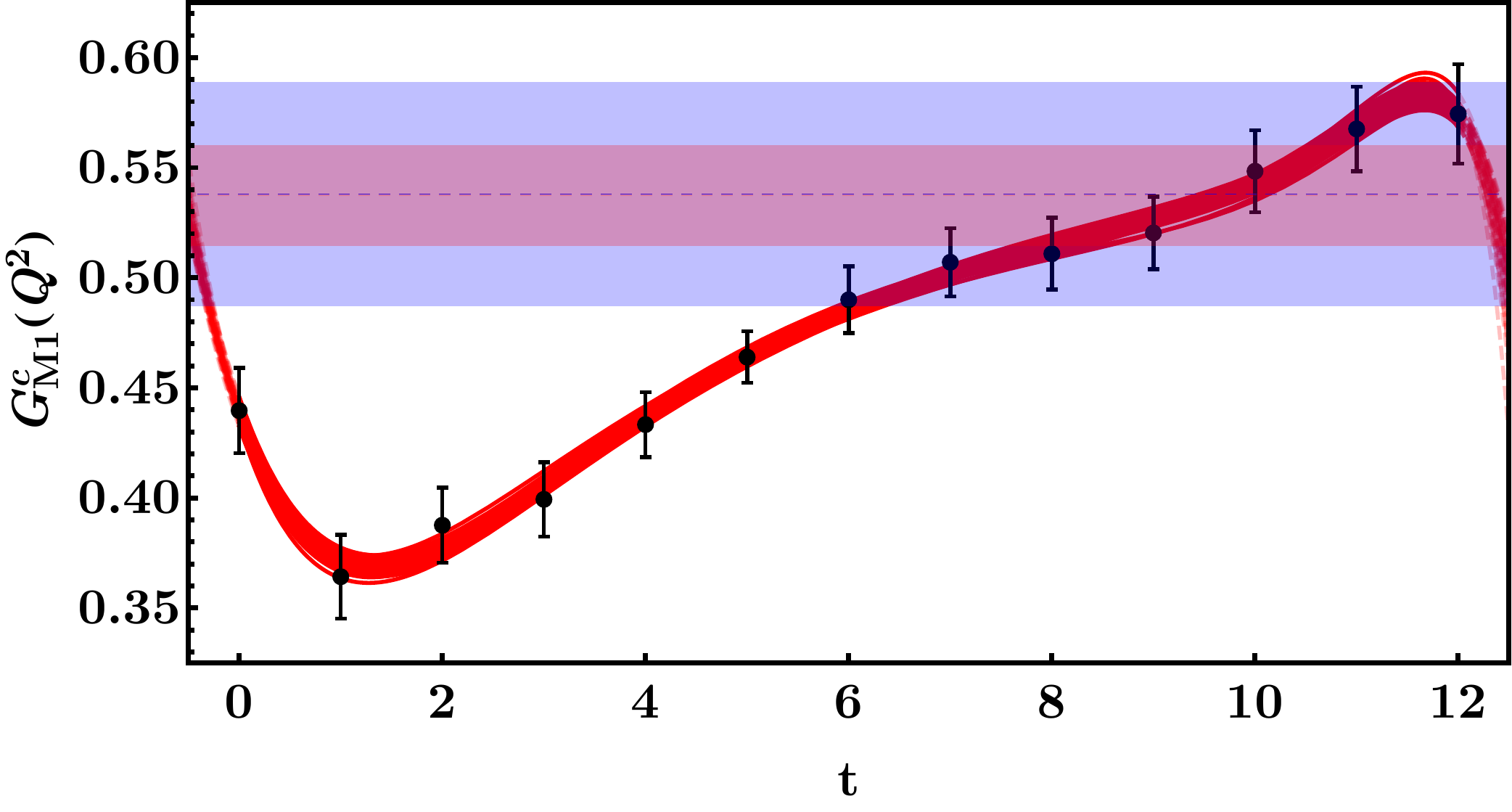}
	\caption{ (Upper) The correlation function ratios $\Pi_1$ and $\Pi_2$ in \Cref{eq:pi1pi2} as functions of current insertion time, $t_1$, for $s$- and $c$-quark sectors of $\Omega_{cc} \gamma \rightarrow \Omega_{cc}^\ast$ transition. $G_{M1}^{s,c}$ obtained via \Cref{eq:mffavg} is also displayed. (Lower) $G_{M1}^{s,c}$ form factors shown with configuration-by-configuration multi-exponential-form fits. Red dashed line with shaded region denotes the weighted average and one standard deviation error of the fit results while blue one is for the average of the results without weighting. Continued dashed curves outside the fit region are there to guide the eye.}
	\label{fig:omega}
\end{figure}

\begin{figure}[htb]
	\includegraphics[width=.49\textwidth]{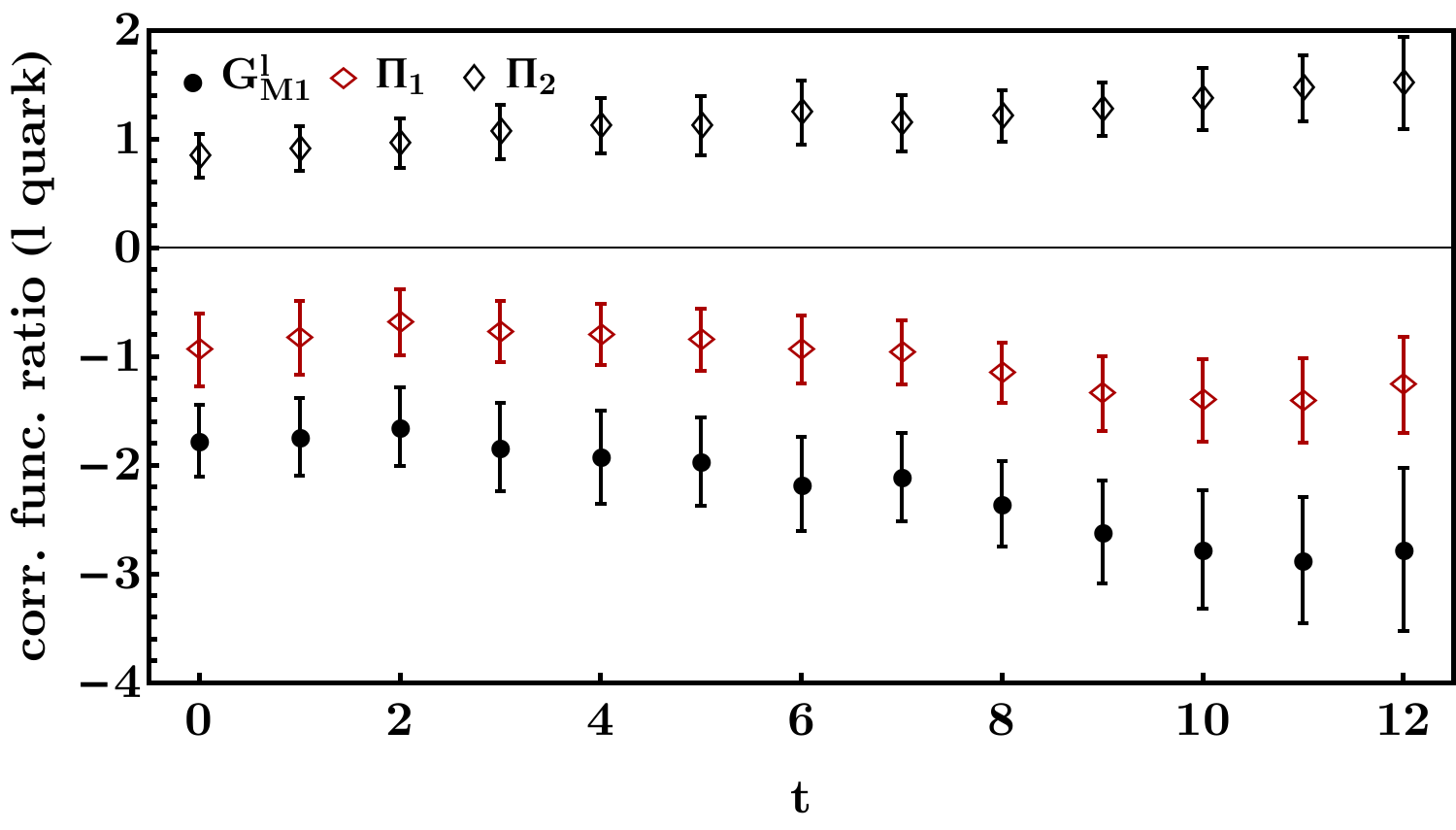} \includegraphics[width=.49\textwidth]{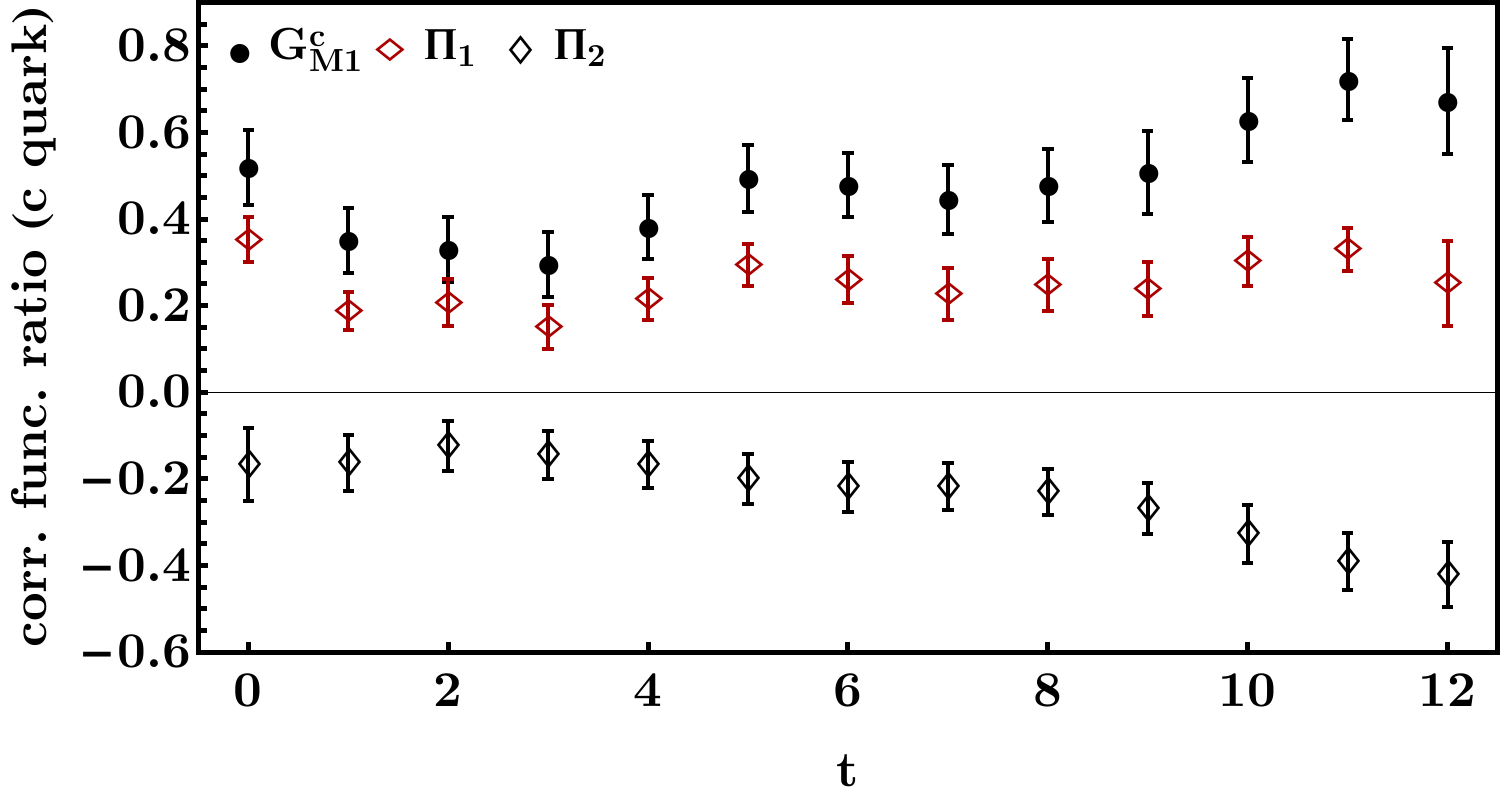} \\
	\includegraphics[width=.49\textwidth]{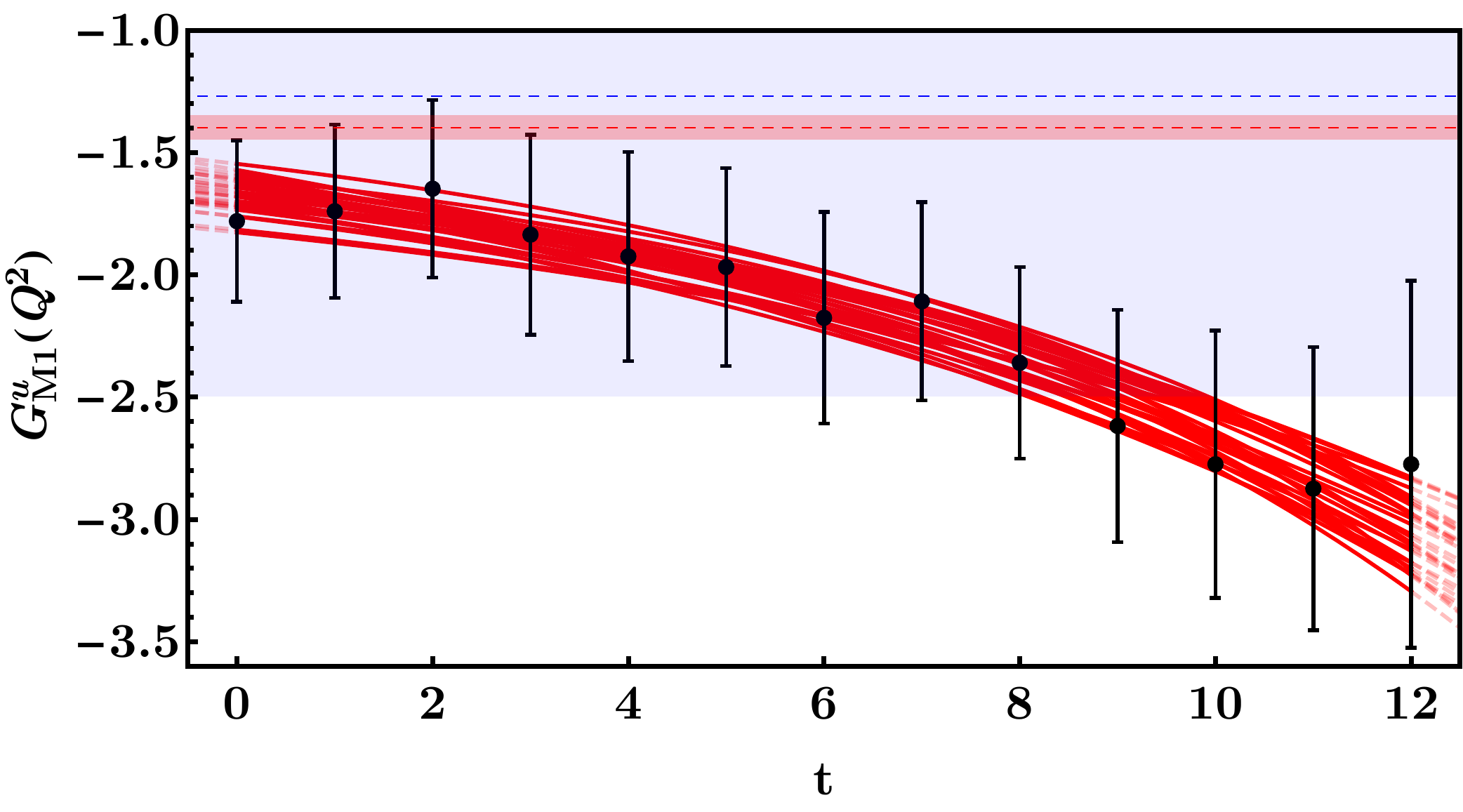} \includegraphics[width=.49\textwidth]{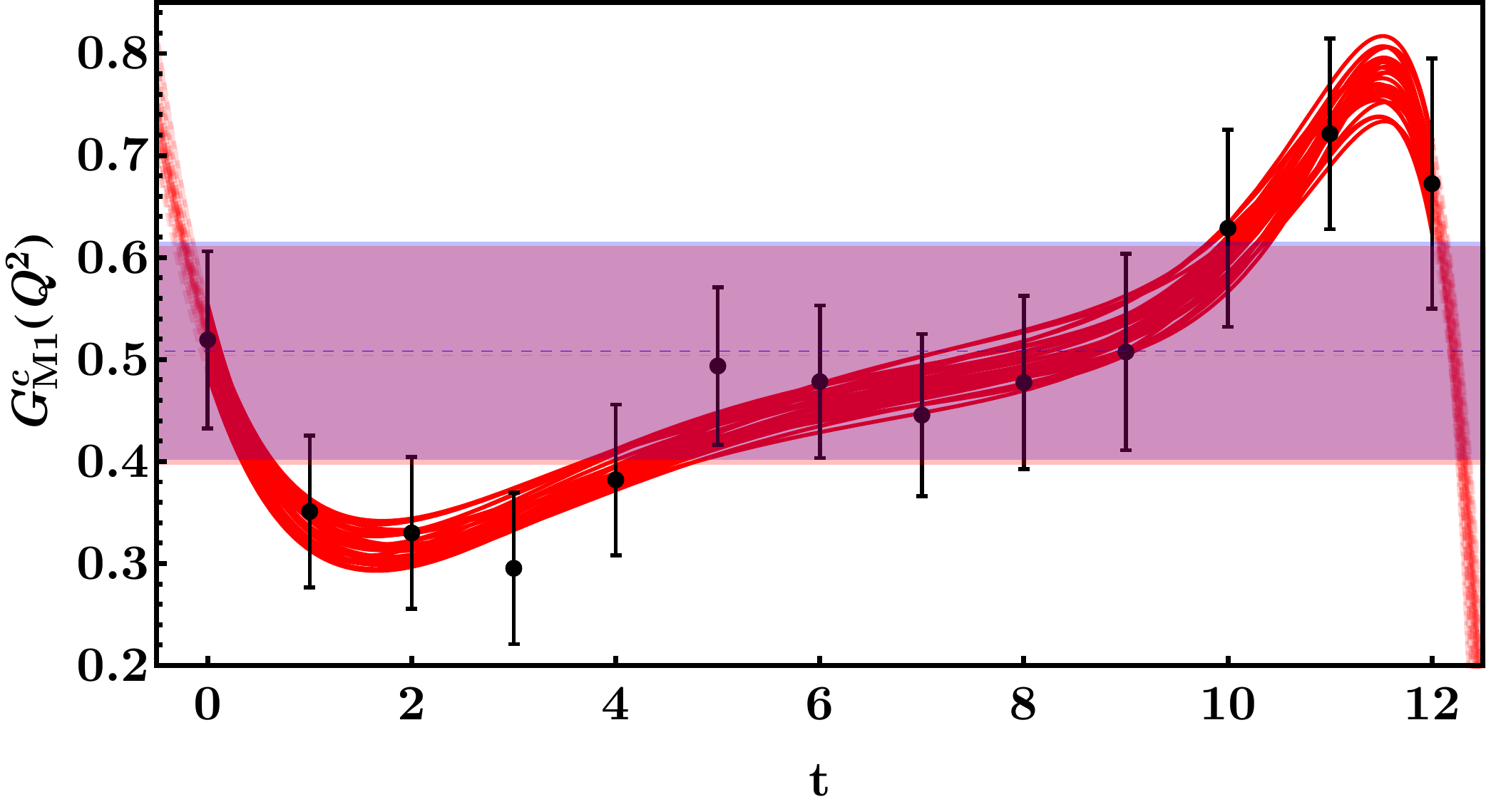}
	\caption{Same as \Cref{fig:omega} but for $\Xi_{cc} \gamma \rightarrow \Xi_{cc}^\ast$ transition. $\ell$ denotes $u$ and $d$ quark for $\Xi_{cc}^{\ast ++}$ and $\Xi_{cc}^{\ast +}$, respectively. }
	\label{fig:xi}
\end{figure}

$\Pi_1$, $\Pi_2$ terms and the $G_{M1}^{(s,\ell),c}(q^2)$ for $\occtooccs$ and $\xcctoxccs$ are illustrated in the upper parts of \Cref{fig:omega,fig:xi} as functions of the current insertion time, $t_1$, for both quark sectors. $\Pi_1$ and $\Pi_2$ contributions have similar magnitudes with opposite signs hence they combine destructively for $G_{E2}$ resulting in a vanishing value. Note that we show the $\Pi_1$ and $\Pi_2$ terms for reference since quoted form factor values are extracted from their proper linear combinations as given in \Cref{eq:mffavg,eq:qffavg}. 
In order to assess the effect of the excited states, we compare the $G_{M1}^{(s,\ell),c}(q^2)$ signal for extended source-sink separations. Our investigations give clear indications that the light and strange quark signals shift significantly, leading us to the conclusion that there are considerable excited-state contaminations that needs to be taken into account. To this end, we consider employing a multi-exponential fit approach to the whole time range of the signal rather than choosing a plateau and performing a constant fit. The general form of the fit function we use is
\begin{equation}\label{eq:multexp}
		R(t_2,t_1) = G_{M1}(q^2) + \sum_i^{N_i} b_i e^{-\Delta_i t_1} + \sum_j^{N_j} b_j e^{-\Delta_j (t_2-t_1)}. 
\end{equation}
First term on the right-hand side corresponds to the form factor value that we want to extract and the following exponentials are there to account for excited-state contributions originating from the source and the sink. $b_i$, $b_j$ and $\Delta_i$, $\Delta_j$ are the overlap factors and mass gaps respectively. Since we have different smearing operators on the source and the sink we leave them as independent free fit parameters. $t_2$ is the fixed sink time slice and $t_1$ is the fit variable current insertion time. We have tried different $N_i=0,1,2,3$ and $N_j=0,1,2,3$ combinations to find the simplest fit function that describes the data. Strange and light quark contributions are contaminated by excited states on the sink side as expected since a wall-smeared operator has a worse overlap to ground state compared to that of Gaussian smeared. We find that two and one exponential from the sink side is enough to represent the excited states for the strange and light quark contributions respectively. Further increasing the number of exponential terms on the sink or adding terms for the source either aggravates the fit quality or yields parameters such that the function can be simplified to the forms that we use. Charm quark contributions on the other hand appear to have a signal that is free from excited state contamination since an $N_i=2$, $N_j=2$ form describes the data with good quality and yields a value that coincides with the data points. 

\begin{figure}[htb]
	\includegraphics[width=.75\textwidth]{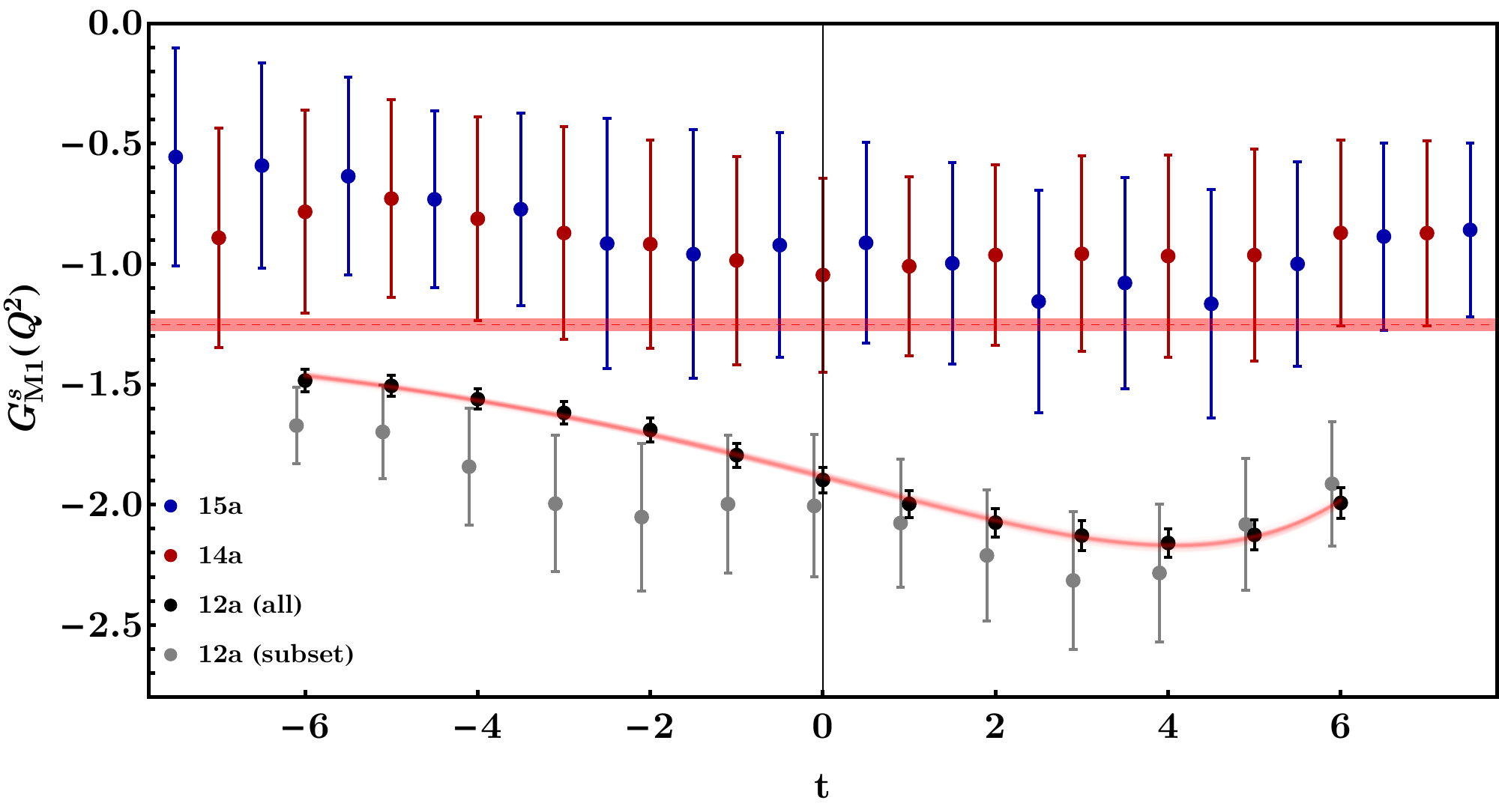}
	\caption{ Comparison of the $G^s_{M1}(q^2)$ signal for extended source-sink separations. $12a$ (subset), $14a$ and $15a$ data points are obtained on a small subset of $44$ configurations while the $12a$ (all) is from the full set of measurements. Points are centered with respect to $t=0$ time slice for the ease of comparison. Red curves and the red shaded region show the multi-exponential-form fits and the weighted average of the fit results. }
	\label{fig:t121415a}
\end{figure}

Multi-exponential fits are illustrated in the lower parts of \Cref{fig:omega,fig:xi}. We take the weighted average of the configuration-by-configuration fit results of $G_{M1}(q^2)$ by considering its parameter error on each configuration as its weight. Red shaded region in \Cref{fig:omega,fig:xi} show the weighted average with 1$\sigma$ deviation while the blue shaded region is for the normal average. Notice that the mean values of the normal and weighted average coincide except for the $\ell$-quark sector of $\Xi_{cc}$, for which, fits on some configurations return poorer results with large parameter errors and averaging without weighting yields a larger deviation. We show the superimposed $G^s_{M1}(q^2)$ signal for extended source-sink separations along with the multi-exponential-form fit result in \Cref{fig:t121415a} to illustrate the excited state analysis. A clear shift in the signal is visible for larger source-sink separations. It is crucial to note that the form factor value we extract via multi-exponential fits agrees nicely with the extended source-sink signals.    

Since the value of $G_{E2}$ is consistent with zero, we do not perform an excited-state analysis, however, it might be more sensitive to other systematic errors. For one, we extract $G_{E2}$ by two numerically differing but analytically identical procedures. First, we compute it by performing fits to the $\Pi_1$ and $\Pi_2$ terms separately and then combining the fit results and secondly, by combining the $\Pi_1$ and $\Pi_2$ terms and then performing a fit to the sum. These two procedures are identical and should result in same values except the numerical fluctuations. We find that these two approaches are consistent with each other. Another source of the systematic error might be due to our omission of the disconnected diagrams. Although their contribution is suppressed with respect to that of connected diagrams, they might become significant since the connected diagram contributions vanish in this case. We expect that the electric quadrupole form factor to be consistent with zero, the reason for the high error for $G_{E2}$ is due to fluctuations of data between negative and positive axis. We observed that the mean values and the standard deviation are slightly changed in further calculations made without using the $G_{E2}$.

\subsubsection{Results}\label{sec:res}
Total form factors can easily be obtained using the individual quark contributions according to the formula,
\begin{equation}\label{eq:individual}
		G_{M1,E2} (Q^2) = \frac{2}{3} G_{M1,E2}^c (Q^2) + c_\ell G_{M1,E2}^\ell (Q^2),
\end{equation}
where $c_\ell = -1/3$ for the $d$  and $s$ quarks and $c_\ell = 2/3$ for the $u$ quark corresponding to $\Xi_{cc}^+$, $\Omega_{cc}^+$ and $\Xi_{cc}^{++}$ baryons, respectively. We use the scaling assumption in \Cref{eq:scaling} to extract the values of the form factors at $Q^2$ = 0. 

Our results for the $M1$ and $E2$ form factors are compiled in \Cref{tab:formf}. Magnetic dipole ($M1$) transition form factor results are given in units of natural magnetons, $\mu_{\cal{B}} \equiv e/2 m_{\cal{B}}$. Note that the charm quark contributions include a factor of $2$ accounting for the number of valence charm quarks. A close inspection of the quark sector contributions shows that the $M1$ form factors are dominantly determined by the light quarks, in agreement with our expectations based on our previous conclusions~\cite{Can:2013tna,Bahtiyar:2015sga,Bahtiyar:2016dom}. The $\ell$-quark contribution is visibly larger than the $c$-quark contribution. This pattern is also consistent with the hyperon transition form factors~\cite{Leinweber:1992pv}: Heavier quark contribution is systematically smaller than that of the light quarks. Contributions of $s$- and $\ell$-quark sector is similar when switching from a $\Omega_{cc}$ baryon to a $\Xi_{cc}$. The charm quark contribution is also similar and suppressed as well which is in agreement with our previous conclusions~\cite{Can:2013tna,Can:2013zpa}. Note that, for the $G_{M1}$ form factors, the absolute mean value of the $\ell$-quark contribution is larger compared to that of $s$-quark.

Previously, we have calculated magnetic moments and charge radii of charmed baryons on a wide range of pion masses changing from $m_\pi \sim 156$ MeV to $m_\pi \sim 700$ MeV~\cite{Can:2013tna,Can:2013zpa,Can:2015exa}. We argue in Ref.~\cite{Can:2015exa} that the finite size effects that might be arising due to $m_\pi L < 4$, are not severe, which we expect to be the case in this calculation too. Moreover, the magnetic moments and the charge radii of the $\Xi_{cc}$ and $\Omega_{cc}$ baryons were found to be similar. Interestingly, magnetic moments of the individual $s$- and $\ell$-quark sectors for $\Xi_{cc}$ and $\Omega_{cc}$ baryons as well were found to be similar within their error bars. Both observations are consistent with the pattern that we see in our current results of $G_{M1}$ form factors of $\xcctoxccsP$ and $\occtooccs$ transitions.

\begin{table}[htb]
\centering
\caption{ Results for $G_{M1}$ and $G_{E2}$ form factors at the lowest allowed four-momentum transfer and at zero momentum transfer. Quark sector contributions to each form factor are given separately weighted with number of valance quarks. $G_{M1}$ results are given in units of natural magnetons ($\mu_{\cal{B}} \equiv e/2 m_{\cal{B}}$).}
\setlength{\extrarowheight}{3pt}
	\begin{tabular*}{\textwidth}{@{\extracolsep{\fill}}cccccccc}
	\hline\hline
									& $Q^2$[GeV$^2$] & $G_{M1}^{\ell}(Q^2)$ & $G_{M1}^{c}(Q^2)$ & $G_{M1}(Q^2)$ & $G_{E2}^{\ell}(Q^2)$ & $G_{E2}^{c}(Q^2)$ & $G_{E2}(Q^2)$ \\
	\hline
	\multirow{2}{*}{$\occtooccs$} 	& 0.181 & -1.252(27)		& 0.537(23)		&  0.775(24)	& -0.034(30)    & 0.002(13) 	& 0.013(14)		\\
	 								& 0     & -1.504(32)  		& 0.571(24)		&  0.882(27) 	& -0.040(36)    & 0.003(14) 	& 0.015(16) 	\\
	\multirow{2}{*}{$\xcctoxccsP$}	& 0.180 & -1.398(50) 		& 0.504(107)	&  0.774(94)	& 0.069(301)   	& -0.005(71)	& -0.026(108) 	\\
	 								& 0     & -1.763(64)	    & 0.528(112)	&  0.906(103)	& 0.087(380)   	& -0.006(75)	& -0.033(133) 	\\
	\multirow{2}{*}{$\xcctoxccsPP$}	& 0.180 & -1.398(50) 		& 0.504(107)	&  -0.552(113)	& 0.069(301)   	& -0.005(71)	& 0.043(210) 	\\
	 								& 0     & -1.763(64)	    & 0.528(112)	&  -0.772(127)	& 0.087(380)   	& -0.006(75)	& 0.054(269) 	\\
	\hline\hline 
	\end{tabular*}
\label{tab:formf}
\end{table}

Sachs form factors can be related to phenomenological observables such as the helicity amplitudes and the decay width of a particle. Relation between the Sachs form factors of a $\cal{B}^\ast$ at rest and the standard definitions of electromagnetic transition amplitudes $f_{M1}$ and $f_{E2}$ are given as~\cite{Nozawa:1989pu, Sato:2000jf}
\begin{align}\label{eq:tamp}
	f_{M1}(q^2)&=\frac{\sqrt{4\pi\alpha}}{2m_{\cal{B}}}\left(\frac{|{\bm{q}}|m_{\cal{B}^\ast}}{m_{\cal{B}}}\right)^{1/2}\frac{G_{M1}(q^2)}{[1-q^2/(m_{\cal{B}}+m_{\cal{B}^\ast})^2]^{1/2}},\\
	f_{E2}(q^2)&=\frac{\sqrt{4\pi\alpha}}{2m_{\cal{B}}}\left(\frac{|{\bm{q}}|m_{\cal{B}^\ast}}{m_{\cal{B}}}\right)^{1/2}\frac{G_{E2}(q^2)}{[1-q^2/(m_{\cal{B}}+m_{\cal{B}^\ast})^2]^{1/2}},
\end{align}
where $\alpha$ is the fine structure constant. Helicity amplitudes $A_{1/2}$ and $A_{3/2}$ are defined as linear combinations of the transition amplitudes as 
\begin{align}\label{eq:hamp}
	A_{1/2}(q^2)&=-\frac{1}{2}[f_{M1}(q^2)+3f_{E2}(q^2)],\\
	A_{3/2}(q^2)&=-\frac{\sqrt{3}}{2}[f_{M1}(q^2)-f_{E2}(q^2)].
\end{align}
The decay width is defined as~\cite{Patrignani:2016xqp}
\begin{equation}
	\Gamma=\frac{ m_{\cal{B}^\ast} m_{\cal{B}}}{8\pi}\left(1-\frac{m_{\cal{B}}^2}{ m_{\cal{B}^\ast}^2}\right)^2\{|A_{1/2}(0)|^2+|A_{3/2}(0)|^2\},
\end{equation} 
in terms of the helicity amplitudes where we have used the constraint ${\bf q}=( m_{\cal{B}^\ast}^2-m_{\cal{B}}^2)/2 m_{\cal{B}^\ast}$ at $q^2=0$. An alternative definition of the decay width in terms of the Sachs form factors can be written as 
\begin{equation}\label{eq:dwG}
	\Gamma=\frac{\alpha}{16}\frac{( m_{\cal{B}^\ast}^2-m_{\cal{B}}^2)^3}{m_{\cal{B}}^2  m_{\cal{B}^\ast}^3}\{3 |G_{E2}(0)|^2+|G_{M1}(0)|^2\}.
\end{equation}
We give our estimates for the helicity amplitudes, decay widths and lifetimes in \Cref{tab:amplitudes}. Both definitions of the decay width give consistent results. Since mass splittings between these baryons kinematically forbid an on-shell strong decay channel, the total decay rates are almost entirely determined in terms of the electromagnetic mode. In comparison to $N \gamma \to\Delta$ transition~\cite{Patrignani:2016xqp}, we observe roughly two order of magnitude suppression in the helicity amplitudes. Considering that the form factors are directly related to the transition matrix elements and thus to the interesting internal dynamics, it is desirable to compare the form factors as well. One can derive the dominant $M1$ form factor of the $N \gamma \to\Delta$ transition by inserting the PDG quoted $A_{1/2}$ and $A_{3/2}$ helicity amplitudes into \Cref{eq:hamp} and following the calculation steps backwards. This calculation returns $G^{M1}_{N\gamma\to\Delta}(0) = 3.063^{+0.102}_{-0.096}$, which is approximately four times greater than the $M1$ form factors of the $\Omega_{cc}^\ast$ and $\Xi_{cc}^\ast$ transitions. Assuming the $u$- and $d$-quark have the same contribution within the $\Delta^+$ baryon, individual quark contributions (without electric charge and quark number factors) can be deduced as $G^{M1,u}_{N\gamma\to\Delta}(0) = G^{M1,d}_{N\gamma\to\Delta}(0) = G^{M1}_{N\gamma\to\Delta}(0)$ with the help of \Cref{eq:individual}. In contrast to the charm quark contributions, this reveals a suppression of around one order of magnitude in $G_{M1}^{c}(0)$. Decay widths are smaller by almost four orders of magnitude, three orders of which are directly related to the similar decrease in the kinematical factor of \Cref{eq:dwG}. $\Omega_{cc}^\ast$, $\Xi_{cc}^{\ast+}$ and $\Xi_{cc}^{\ast++}$ have similar decay widths and lifetimes.
\begin{table}[htb]
\centering
\caption{Results for the helicity amplitudes, decay widths and lifetimes. Zero-momentum values are obtained using the simple scaling assumption given in \Cref{eq:scaling}.}
\setlength{\extrarowheight}{3pt}
	\begin{tabular*}{\textwidth}{@{\extracolsep{\fill}}cccccccc}
	\hline\hline 
	& $Q^2$ & $f_{M1}$ & $f_{E2}$ & $A_{1/2}$ & $A_{3/2}$ & $\Gamma$ & $\tau$ \\   
	& [GeV$^2$]	& \scriptsize{$10^{-2}$[GeV$^{-1/2}$]} & \scriptsize{$10^{-2}$[GeV$^{-1/2}$]} & \scriptsize{$10^{-2}$[GeV$^{-1/2}$]} & \scriptsize{$10^{-2}$[GeV$^{-1/2}$]} & \scriptsize{[keV]}& \scriptsize{[$10^{-18}$ s]} \\
	\hline 
	\multirow{2}{*}{$\occtooccs$} 	& 0.181 & 0.812(26) 	& 0.013(15)	    & -0.429(13)	& -0.690(22)	& --- 			& --- 			\\
	 								& 0     & 0.924(28)  	& 0.016(17)	    & -0.489(14) 	& -0.785(25) 	& 0.0565(4)		& 11.66(3.83) 	\\
	\multirow{2}{*}{$\xcctoxccsP$}  & 0.180 & 0.838(101) 	& -0.027(118) 	& -0.419(51)  	& -0.726(88) 	& --- 			& --- 			\\
	 								& 0     & 0.982(111)  	& -0.034(145)  	& -0.491(56)  	& -0.850(96)  	& 0.0648(38) 	& 10.28(3.30) 	\\
	\multirow{2}{*}{$\xcctoxccsPP$} & 0.180 & -0.597(123)	& 0.048(229) 	& 0.298(61) 	& 0.517(106)  	& --- 			& --- 		    \\
									& 0     & -0.835(137) 	& 0.061(293) 	& 0.417(69) 	& 0.723(119)    & 0.0518(56) 	& 12.70(2.04) 	\\
	\hline\hline
	\end{tabular*}
\label{tab:amplitudes}
\end{table}

\subsubsection{Comparison to non-lattice methods}\label{sec:comp}
Electromagnetic transitions of the doubly charmed baryons have also been studied within the heavy hadron chiral perturbation theory \cite{Meng:2017dni,Li:2017pxa,Li:2017cfz}, covariant baryon chiral perturbation theory \cite{Liu:2018euh}, bag model \cite{Hackman:1977am,Bernotas:2013eia}, quark models \cite{SilvestreBrac:1996bg,Lichtenberg:1976fi,JuliaDiaz:2004vh,Faessler:2006ft,Branz:2010pq,Oh:1991ws} and QCD sum rules \cite{Sharma:2010vv}. Electromagnetic decays of doubly charmed baryons are found to be suppressed, which is qualitatively in agreement with our results. Bag model predictions~\cite{Hackman:1977am,Bernotas:2013eia} for decay widths are one order of magnitude larger than our results. Quark model predictions are even larger by two orders of magnitude \cite{Xiao:2017udy,Lu:2017meb,Branz:2010pq} similar to those of the chiral perturbation theory \cite{Li:2017pxa} and QCD sum rules \cite{Cui:2017udv}. In order to understand the discrepancy between our and non-lattice results, we compile the masses and the decay widths of various non-lattice methods as well as the calculated mass splittings, kinematic factors and $M1$ form factor values relevant to the $\occtooccs$ transition in \Cref{tab:nonlat} for comparison. Kinematic factor ($K.F.$) is $( m_{\cal{B}^\ast}^2-m_{\cal{B}}^2)^3/m_{\cal{B}}^2  m_{\cal{B}^\ast}^3$ in \Cref{eq:dwG}. 

\begin{table}[htb]
\centering
\caption{Comparison to non-lattice methods. We calculate the mass splittings, kinematic factors ($K.F.$) and $M1$ form factor values of other methods by inserting their respective mass and decay width values.}
\setlength{\extrarowheight}{3pt}
	\begin{tabular*}{\textwidth}{@{\extracolsep{\fill}}c|c|ccccccc}
	\hline\hline 
    & This work & Ref. \cite{Hackman:1977am} & Ref. \cite{Bernotas:2013eia} & Ref. \cite{Lu:2017meb} & Ref. \cite{Branz:2010pq} & Ref. \cite{Xiao:2017udy} & Ref. \cite{Li:2017pxa} & Ref. \cite{Cui:2017udv} \\ \hline	 
\multicolumn{1}{r|}{$m_{\Omega_{cc}}$ [GeV]}     	& 3.719(10) & 3.781 & 3.815 & 3.715 & 3.778 & 3.778 & 3.620 & 3.778 \\ 
\multicolumn{1}{r|}{$m_{\Omega_{cc}^\ast}$ [GeV]}	& 3.788(11) & 3.854 & 3.876	& 3.772 & 3.872 & 3.872 & 3.720 & 3.872 \\ 
\multicolumn{1}{r|}{$m_{\Omega_{cc}^\ast}-m_{\Omega_{cc}^\ast}$ [MeV]}	& 69 & 73 & 61 & 57 & 94 & 94 & 100 & 94 \\ 
\multicolumn{1}{r|}{$\Gamma(\occtooccs)$ [keV]} 	& 0.0565(4) & 1.35 & 0.949 & 0.82 & 2.11(11) & 6.93	& 9.45 & 5.4$^{+6.9}_{-3.1}$ \\
\hline
\multicolumn{1}{r|}{$(K.F.)_{\Omega_{cc}}$ $\times 10^{-3}$ [GeV]} & 0.185 & 0.212 & 0.122 & 0.105 & 0.449 & 0.449	& 0.586 & 0.449 \\ 
\multicolumn{1}{r|}{$G_{M1}^{\occtooccs}$ [$\mu_{\cal{B}}$]} 			& 0.882(27) & 3.739 & 4.132 & 4.139 & 3.210(732) & 5.818 & 5.945 & 5.136$^{+5.389}_{-3.891}$ \\
 \hline \hline
\end{tabular*}
\label{tab:nonlat}
\end{table}

As we have discussed in \Cref{sec:res}, the decay widths of the transitions that we consider in this work are narrower mainly due to the decrease in the kinematic factors in contrast to that of the $N\gamma\to\Delta$ transition. Comparison of the kinematic factors suggests that the discrepancy with the non-lattice methods arises from the $M1$ form factors. $G_{M1}$ values of the non-lattice methods are close to or larger than the $N\gamma\to\Delta$ value, which is highly unlikely since we find that the heavy-quark contribution to $M1$ transition is heavily suppressed and the light quark contribution is not enhanced enough to compensate the change. $E2$ transitions, on the other hand, almost vanish so that they do not play a significant role. Although it is plausible that there may be uncontrolled systematic errors affecting our results we remind the reader that \emph{\textbf{i)}} our results are free from chiral extrapolation errors since the ensembles we use are almost at the physical-quark point, \emph{\textbf{ii)}} any discretization error arising from the charm-quark action is suppressed and controlled since we employ a relativistic heavy quark action, \emph{\textbf{iii)}} we have identified and included the effect of the excited-state contamination in our analysis and \emph{\textbf{iv)}} based on our analysis in Ref. \cite{Can:2015exa}, we expect the finite size effects on these configurations to be less than $1\%$. Systematics that might arise from continuum extrapolation, however, remains unchecked. It is intriguing that we have observed a similar, but less drastic discrepancy, in $M1$ form factors (or magnetic moments) in our previous works of diagonal spin-$1/2 \to$ spin-$1/2$ transitions where our results \cite{Can:2013tna,Bahtiyar:2016dom} are smaller compared to that of model estimations. Discrepancies between lattice and non-lattice results are still an issue that needs to be understood better from both sides.  

\subsection{Systematic errors on charm quark observables}\label{sec:se}  
Since we switch to a relativistic heavy quark action in this analysis, while keeping the rest of the setup the same, we use this opportunity to quantify the systematic errors on charm observables in comparison to using a Clover action prescription~\cite{Bahtiyar:2015sga}. To this end, we re-calculate the $\Omega_c \gamma \rightarrow \Omega_c^\ast$ transition form factors, which follows the same procedures described in previous sections. Note that we use plateau method in this case to extract the form factors since extended source-sink separation and $12a$ signals coincide. A comparison of our results are given in \Cref{tab:comparison}. Note that the $\kappa_s^{\text{val}}$ value we use in this and the previous work differs, therefore the change in $\Omega_c$ and $\Omega_c^\ast$ masses cannot solely be attributed to the change of the charm quark action. Strange quark observables also differ due to the same reason. $G_{E2}^{c}(Q^2)$ is not a reliable observable either since its charmed-sector results are consistent with zero in both cases. A clear comparison can be made using the $G_{M1}^{c}(Q^2)$ form factor for which we see a $\sim$ 20\% deviation. We provide the full results of the analysis from $730$ measurements in \Cref{tab:formf_oc,tab:amplitudes_oc} for completeness. The updated decay width is $\Gamma = 0.096(14)$ keV, approximately 20\% larger than but still in agreement within errors with the previous estimation of $\Gamma = 0.074(8)$ keV~\cite{Bahtiyar:2015sga}, leaving the conclusions unchanged.

\begin{table}[!htb]
\centering
\caption{Mass of $\Omega_c$ and $\Omega_c^\ast$ as well as the charmed-sector of the $\Omega_c \gamma \rightarrow \Omega_c^\ast$ transition form factors at $Q^2=0.180$ GeV$^2$.}
\label{tab:comparison}
\setlength{\extrarowheight}{3pt}
	\begin{tabular*}{\textwidth}{@{\extracolsep{\fill}}lcccc}
	\hline\hline
		& $m_{\Omega_c}$ [GeV] 	& $m_{\Omega_c^\ast}$ [GeV] & $G_{M1}^{c}(Q^2)$ [$\mu_{\cal{B}}$] & $G_{E2}^{c}(Q^2)$\\
	\hline
	Bahtiyar et al.~\cite{Bahtiyar:2015sga}	& $2.750(15)$ & $2.828(15)$	& $-0.167(33)$ & $-0.008(26)$ \\
	This work 								& $2.707(11)$ & $2.798(24)$	& $-0.209(30)$ & $-0.010(23)$ \\
	Exp.									& $2.695(2)$  & $2.766(2)$  & --- & --- \\
	\hline\hline
	\end{tabular*}
\end{table}

\begin{table}[htb]
\centering
\caption{ Results for $G_{M1}$ and $G_{E2}$ form factors of the $\Omega_c \gamma \rightarrow \Omega_c^\ast$ transition at the lowest allowed four-momentum transfer and at zero momentum transfer. Quark sector contributions to each form factor are given separately. $G_{M1}$ results are given in units of natural magnetons, $\mu_{\cal{B}}$.}
\setlength{\extrarowheight}{3pt}
\begin{tabular*}{\textwidth}{@{\extracolsep{\fill}}ccccccc}
	\hline\hline 
	$Q^2$[GeV$^2$] & $G^s_{M1}(Q^2)$ & $G^c_{M1}(Q^2)$ & $G_{M1}(Q^2)$ & $G^s_{E2}(Q^2)$ & $G^c_{E2}(Q^2)$ & $G_{E2}(Q^2)$  \\
	\hline
	0.180 & $1.456(102)$ & $-0.209(30)$ & $-0.625(43)$  & $-0.195(11)$ & $0.010(23)$ & $0.059(43)$ \\
	0     & $1.748(122)$ & $-0.215(31)$ & $-0.725(50)$  & $-0.234(134)$ & $0.010(24)$ & $0.071(52)$ \\
	\hline \hline	
\end{tabular*}
\label{tab:formf_oc}
\end{table}

\begin{table}[htb]
\centering
\caption{Results for the helicity amplitudes and the decay width of the $\Omega_c \gamma \rightarrow \Omega_c^\ast$ transition. Helicity amplitudes are given at finite and zero momentum transfer. Zero-momentum values are obtained using the scaling assumption in \Cref{eq:scaling}.}
\setlength{\extrarowheight}{3pt}
\begin{tabular*}{\textwidth}{@{\extracolsep{\fill}}ccccccccc}
	\hline\hline 
	$Q^2$ & $f_{M1}$ & $f_{E2}$ & $A_{1/2}$ & $A_{3/2}$ & $\Gamma$ & $\tau$ \\
	\scriptsize{[GeV$^2$]}	& \scriptsize{$10^{-2}$[GeV$^{-1/2}$]} & \scriptsize{$10^{-2}$[GeV$^{-1/2}$]} & \scriptsize{$10^{-2}$[GeV$^{-1/2}$]} & \scriptsize{$10^{-2}$[GeV$^{-1/2}$]} & \scriptsize{[keV]} & \scriptsize{[$10^{-18}$ s]} \\
	\hline
	0.180 	& $-0.951(66)$ & $-0.090(65)$ & $0.341(99)$ & $0.901(85)$ & --- & --- \\
	0 		& $-1.104(76)$ & $0.109(79)$  & $0.389(119)$ & $1.050(101)$ & $0.096(14)$ & $6.889(997)$ \\
	\hline\hline
\end{tabular*}
\label{tab:amplitudes_oc}
\end{table}

\section{Summary and Conclusions}\label{sec:sc}
We have evaluated the radiative transitions of doubly charmed baryons in 2+1-flavor lattice QCD and extracted the magnetic dipole ($M1$) and electric quadrupole ($E2$) form factors as well as the helicity amplitudes and the decay widths. We have extracted the individual quark contributions to the $M1$ and $E2$ form factors and found that $M1$ form factors are dominantly determined by the light quarks. $E2$ form factor contributions are found to be negligibly small and its absence has a minimal effect on the observables. The helicity amplitudes are observed to be suppressed roughly by two order of magnitude in comparison to the $N\gamma\to\Delta$ transition's. $M1$ form factors are found to be suppressed by less than an order with respect to the $N\gamma\to\Delta$, suggesting that the kinematical factors play a more important role in suppressing the helicity amplitudes and the decay widths in the heavy quark systems. $\Omega_{cc}^\ast$ and $\Xi_{cc}^\ast$ have roughly the same decay width and lifetime. Our results qualitatively agree with the predictions of other approaches however there is a quantitative disagreement of around one or more than one order of magnitude, which calls for more investigations to resolve. We have also provided updated results for the $\Omega_c \gamma \rightarrow \Omega_c^\ast$ transition computed with a relativistic heavy quark action and estimated the systematic error due to using a Clover action. Our results are particularly suggestive for experimental facilities such as LHCb, PANDA, Belle II and BESIII to search for further states.

\acknowledgments
The unquenched gauge configurations employed in our analysis were generated by PACS-CS collaboration~\cite{Aoki:2008sm}. We used a modified version of Chroma software system~\cite{Edwards:2004sx} along with QUDA~\cite{Babich:2011np,Clark:2009wm}. K. U. Can thanks Dr. Balint Joo for his guidance on the Chroma software system and Dr. Yusuke Namekawa for discussions on Tsukuba action and his comments on the manuscript. This work is supported in part by The Scientific and Technological Research Council of Turkey (TUBITAK) under project number 114F261 and in part by KAKENHI under Contract Nos. 25247036 and 16K05365.

\end{document}